\documentclass[12pt]{article}

\usepackage{scicite}

\usepackage{times}

\usepackage{graphicx}

\usepackage{xcolor}

\usepackage{amsmath}

\usepackage{caption}

\usepackage{newfloat}

\DeclareFloatingEnvironment{Supplementary Figure}

\newenvironment{sciabstract}{%
\begin{quote} \bf}
{\end{quote}}

\title{Detecting the spin-polarization of edge states in graphene nanoribbons}

\author
{Jens Brede,$^{1,2}$ Nestor Merino-D\'{i}ez,$^{1,2}$ Alejandro Berdonces,$^{1,2}$ Sof\'{i}a Sanz,$^{1}$\\
Amelia Dom\'{i}nguez-Celorrio,$^{3}$ Jorge Lobo-Checa,$^{3,4,5}$ Manuel Vilas-Varela,$^{6}$\\ Diego Pe\~{n}a,$^{6}$ Thomas Frederiksen,$^{1,7}$ Jose I. Pascual,$^{7,8\ast}$\\ Dimas G. de Oteyza,$^{1,2,9\ast}$ David Serrate$^{3,4,5\ast}$\\
\\
\normalsize{$^{1}$Donostia International Physics Center, E-20018 San Sebasti\'an, Spain}\\
\normalsize{$^{2}$Centro de F\'isica de Materiales (MPC) CSIC-UPV/EHU, E-20018 San Sebasti\'an, Spain}\\
\normalsize{$^{3}$Instituto de Nanociencia y Materiales de Arag\'on (INMA), CSIC-Universidad de Zaragoza,}\\
\normalsize{E-50009 Zaragoza, Spain}\\
\normalsize{$^{4}$Departamento de Física de la Materia Condensada, Universidad de Zaragoza,}\\\normalsize{E-50009, Zaragoza, Spain}\\
\normalsize{$^{5}$Laboratorio de Microscopias Avanzadas (LMA), Universidad de Zaragoza,}\\\normalsize{ E-50018, Zaragoza, Spain}\\
\normalsize{$^{6}$Centro Singular de Investigaci\'on en Qu\'imica Biol\'oxica e Materiais Moleculares (CiQUS)}\\
\normalsize{and Departamento de Qu\'imica Org\'anica, Universidade de Santiago de Compostela;}\\
\normalsize{E-15782 Santiago de Compostela, Spain}\\
\normalsize{$^{7}$Ikerbasque, Basque Foundation for Science, E-48013 Bilbao, Spain}\\
\normalsize{$^{8}$CIC nanoGUNE BRTA, E-20018 San Sebasti\'an, Spain}\\
\normalsize{$^{9}$Nanomaterials and Nanotechnology Research Center (CINN), CSIC-UNIOVI-PA}\\
\normalsize{E-33940 El Entrego, Spain}\\
\normalsize{$^\ast$To whom correspondence should be addressed; E-mail:  serrate@unizar.es,}\\
\normalsize{d.g.oteyza@cinn.es, ji.pascual@nanogune.eu}
}

\date{}

\begin{document}


\baselineskip24pt


\maketitle


\begin{sciabstract}
      Low dimensional carbon-based materials are interesting because they can show intrinsic $\pi$-magnetism associated to p-electrons residing in specific open-shell configurations. Consequently, during the last years there have been impressive advances in the field combining indirect experimental fingerprints of localized magnetic moments with theoretical models. In spite of that, a characterization of their spatial- and energy-resolved spin-moment has so far remained elusive. To obtain this information, we present an approach based on the stabilization of the magnetization of $\pi$-orbitals by virtue of a supporting substrate with ferromagnetic ground state. Remarkably, we go beyond localized magnetic moments in radical or faulty carbon sites: In our study, energy-dependent spin-moment distributions have been extracted from spatially extended one-dimensional edge states of chiral graphene nanoribbons. This method can be generalized to other nanographene structures,  representing an essential validation of these materials for their use in spintronics and quantum technologies.
\end{sciabstract}


Magnetism is at the heart of a vast amount of applications and conventionally relies on the spin of unpaired $d$- or $f$-shell electrons. Instead, carbon-based magnetism stems from $p$-shell electrons.\cite{Yazyev_Emergence_2010,de_oteyza_carbon_2022} The vision of exploiting them in new types of applications involving spin-polarized currents and spin-based quantum information\cite{Yazyev_Emergence_2010, de_oteyza_carbon_2022, Slota_Magnetic_2018, Lombardi_Quantum_2019, Trauzettel_SpinQubits_2007} is inspired in two distinct properties: weak spin-orbit and hyperfine coupling (two of the main channels responsible for the relaxation and decoherence of electron spins),\cite{Yazyev_Emergence_2010, de_oteyza_carbon_2022, Min_Intrinsic_2006, Yazyev_Hyperfine_2008, Slota_Magnetic_2018} and delocalization in $\pi$-orbitals with high spin-wave stiffness.\cite{Yazyev_Emergence_2010, de_oteyza_carbon_2022, yazyev_magnetic_2008} However, the experimental realization of the associated open-shell molecular structures is challenging. Therefore, it is only recently that they are becoming accessible by on-surface synthesis under vacuum conditions,\cite{Song_On-Surface_2021}  rendering their characterization of utmost interest.

Within the wide range of carbon nanostructures predicted to display  $\pi$-magnetism,\cite{Song_On-Surface_2021} graphene nanoribbons (GNRs, i.e., one-dimensional stripes of graphene) are probably among the most interesting for potential applications due to their intrinsic length,\cite{Corso_bottom-up_2018, saraswat_materials_2021} which facilitates contacting and integration into device structures.\cite{Bennet_Bottom-up_2013, Llinas_Short_2017} Aside from the presence of vacancies or heteroatoms, for GNRs to exhibit magnetic properties they must display zigzag edges, whether continuously throughout the ribbon (zGNRs) or in periodic alternation with armchair segments (chiral or chGNRs).\cite{Corso_bottom-up_2018, yazyev_theory_2011, carvalho_edge_2014} In either case, the ribbons develop edge states that decay exponentially towards the ribbon's interior. These electronic states are predicted to be spin-polarized.\cite{Corso_bottom-up_2018,yazyev_theory_2011, carvalho_edge_2014} Both, zGNRs and chGNRs, have been synthesized with atomic precision on Au(111) substrates.\cite{ruffieux_-surface_2016,de_oteyza_substrate-independent_2016, merino-diez_unraveling_2018, li_topological_2021} Whereas the presence of the edge states has been confirmed for both cases,\cite{ruffieux_-surface_2016, li_topological_2021}, a direct experimental proof of their spin polarization is still lacking. In the more general case of open-shell carbon nanostructures, spatially resolved magnetic signals have only been observed in a few cases involving magnetic field-dependent Zeeman splittings\cite{li_uncovering_2020, mishra_topological_2020, Zheng_Engineering_2020}. Otherwise, only indirect hints of the magnetism have been obtained, whether from an analysis of the frontier orbital's density of states\cite{pavlicek_synthesis_2017,blackwell_spin_2021}, Kondo resonances,\cite{li_single_2019, li_uncovering_2020, mishra_topological_2020} inelastic spin-flip excitations\cite{li_single_2019,Zheng_Engineering_2020} or Coulomb gaps.\cite{wang_giant_2016, ruffieux_-surface_2016} In any case, the detection of an intrinsic remanent spin-polarization of $\pi$-orbitals has never been obtained to date for any carbon-based material. Actually, the weak spin-orbit coupling\cite{Slota_Magnetic_2018}, a central advantage of carbon-based magnetism, imposes at the same time the main drawback to resolve a stationary spin moment in these systems: the practically null magnetic anisotropy of $sp_2$ carbon atoms.

To circumvent this constrain, in this work we utilize a GdAu$_2$ ferromagnetic monolayer on Au(111) to unambiguously demonstrate the spin-polarization of chGNR edge states atop it.\cite{corso2010_GdAu2} We obtain chGNRs with edges oriented along the chiral (3,1) graphene lattice vector\cite{yazyev_theory_2011} and $c=8$ carbon atoms across their width (thus (3,1,8)-chGNRs) by deposition and appropriate thermal treatments (see Methods) of the reactant 2”,3’-dibromo-9,9’:10’,9”:10”,9”’-quateranthracene \cite{li_topological_2021} (DBQA). Subsequent characterization by spin-polarized scanning tunneling microscopy and spectroscopy (SP-STM/STS), supported by mean-field Hubbard (MFH) model calculations, unravel the spatially and energetically resolved spin-polarization of the ribbon's frontier states.

For the characterization of an unexplored magnetic ground state by means of SP-STM, it is convenient to arrange the sample under study in coexistence with a surface that has a well known spin-resolved electronic structure. A GdAu$_2$ monolayer on Au(111)\cite{corso2010_GdAu2} is an excellent candidate. Its ability to catalyze nanographene polymerization via Ullmann coupling\cite{talirz2016} has been already proven\cite{abadia_polymerization_2017,que_-surface_2020} and, at the same time, it orders ferromagnetically below 19 K\cite{cavallin2014} with a large easy-plane magnetic anisotropy, thereby showing strong in-plane contrast in SP-STM measurements\cite{bazarnik19}.

We achieve long (3,1,8)-chGNRs from DBQA precursor molecules (Supplementary Fig. S1) on GdAu$_2$ using a very similar procedure as the one previously reported in the case of Au(111)\cite{li_topological_2021} (see Supplementary Experimental Methods). Fig. 1A shows an overview image of (3,1,8)-chGNRs on GdAu$_2$. The moir\'{e} superlattice caused by the superposition of the GdAu$_2$ lattice (hexagonal unit cell of 5.41$\pm 0.03$ {\AA}) and the underlying Au(111) lattice\cite{corso2010_GdAu2,corso2010} is clearly visible. In the GdAu$_2$ lattice, each Gd atom is sixfold coordinated with Au atoms, which appear in STM images as dark and bright spots, respectively (Fig. 1B). The edges in (3,1,8)-chGNRs are composed of alternating segments of zigzag and armchair graphene paths (see molecular scheme in Fig. 1B).  The GNRs grow preferentially with their longitudinal axis along high symmetry directions of the Gd atomic lattice, either the $[1\overline{1}0]$ (e.g. the central ribbon in Fig. 1B) or the $[\overline{2}11]$ directions of the Au(111) substrate.

In addition to enabling the formation of high quality (3,1,8)-chGNRs, the GdAu$_2$ surface exhibits different domains of the in-plane magnetization, as revealed in the spin-resolved dI/dV maps at $V_{b}=3$ V shown in Figs. 1C-D, taken with bulk Cr-tips (see also Supplementary Fig. S2)\cite{bazarnik19}. The magnetic coupling across crystallographic antiphase domain boundaries (APB) is such that the magnetization components along the tip sensitivity direction at either side are antiparallel. This provides atomically sharp boundaries between GdAu$_2$ regions with different magnetization (Figs. 1E-F) that serves as a control of the unaltered tip's spin sensitivity throughout the measurement process. We select target GNRs located in a magnetic domain with homogeneous magnetization and close to an APB. Ramping the external out-of-plane field to $\pm3$ Tesla and back to zero allows us to repeatedly switch the remanent state of the substrate underneath the target GNR (see Supplementary Note 2), as shown in Figs. 1D-F and sketched in Fig. 1G. By convention, we refer to magnetic states with high and low differential conductance at 3 V as parallel ($P$) and antiparallel ($AP$) states to the spin direction of the tip.

After setting the magnetic state of the substrate, we proceed with the characterization of the target GNR (Fig. 1H, the characterization region is marked in Fig. 1C, lying on the left domain of Figs. 1E-F). The density of states (DoS) around the Fermi level can be by retrieved from low-bias $dI/dV$ maps. The result for a ribbon with $N=17$ precursor units (Fig. 1H), measured with metallic tips (Cr or W), is representative of DoS obtained for other GNRs with lengths varying from 10 to 23 precursor units. The internal structure observed in the central region corresponds to the GdAu$_2$ lattice, and the moiré contrast is visible through the ribbon.

Importantly, we resolve the predicted high DoS at the ribbons's edges \cite{yazyev_theory_2011,carvalho_edge_2014,li_topological_2021}. The well-known short decay length of edge states into the vacuum \cite{sode_electronic_2015} suppresses their characteristic intensity in constant height DoS images taken with metallic tips (Supplementary Fig. S4). However, the expected DoS distribution is readily recovered when using CO-functionalized W-tips, as shown in Fig. 2A for a $N=21$ chGNR, including the structure of the wave function extending into the central part of the ribbon. The pronounced edge states are, in most cases, further modulated by the moir\'{e} pattern underneath the edge (See Supplementary Note 1), rather than displaying pure quantum well states caused by the finite length\cite{carbonell-sanroma_quantum_2017,merino-diez_unraveling_2018}.

SP-STM experiments require a metallic Cr-tip, and thus we cannot resort to CO functionalization. To recover the sensitivity to the molecular states with this kind of tips it is necessary to acquire grids of $dI/dV$ spectra over the chGNR regulating the current at each pixel (see Methods). Selected point spectra are shown in Fig. 2C, exhibiting some peaks between -6 mV and 28 mV, and a first fully occupied state at -40 mV. Fig. 2D shows $dI/dV$ slices at constant $V_b$ for these states, where a certain repetition period along the edge can be discerned. By fast Fourier transform these periodicities are turned into characteristic wave vectors for each molecular state,\cite{merino-diez_unraveling_2018,sode_electronic_2015} which permits to assign them to discretized states emerging from the conduction and valence bands of the infinitely long GNRs (See Supplementary Fig. S4). The peak at -40 mV arises from the valence band, and thus corresponds to the Highest Occupied Molecular Orbital (HOMO) of the charge-neutral ribbon. The peaks crossing the Fermi level ($V_{b}=0$ mV)  and above ($V_{b}\sim20$ mV) relate to the conduction band and are the Lowest Unoccupied Molecular Orbitals (LUMO, LUMO+1) of the charge-neutral ribbon. The appearance of the LUMO right at the Fermi level on GdAu$_2$/Au(111) is a distinct feature of charge transfer from the substrate to the GNR. This is a consequence of the reduced work function of GdAu$_2$/Au(111) with respect to Au(111)\cite{que_-surface_2020}, which promotes a slight electron doping of the GNR and drives the LUMO states stemming from the bottom of the conduction band below the Fermi level.

The LUMO peaks centered at zero bias exhibit an additional fine splitting of approximately 12 mV  across the Fermi level at some locations (e.g. curves 2 and 3 in Fig. 2C). This peak sub-structure is observed whenever a sizable DoS centered at zero bias is present at chiral edges. To elucidate the origin of the edge electronic structure, we solved the tight binding Hamiltonian for the $\pi$-electron system including a MFH term to account for electronic correlations (see Methods). For non-interacting electrons ($U=0$), the addition of two electrons to the charge-neutral ribbon shifts the simulated DoS spectra at the chiral edge by about -8 meV, (Fig. 2E). The original LUMO+1 at 34 meV sits now at 26 meV, and the zero-energy end state of topological origin\cite{li_topological_2021} located at the termini shifts to -8 meV. We introduce electron-electron (\emph{e-e}) interactions via an on-site Coulomb repulsion term $U\neq0$. With $U=1$ eV we are able to reproduce our experimental $dI/dV$ spectra (top row in Fig. 2E). First, the main intensity of the first fully occupied states clusters around -40 meV. Second, the structure around the Fermi level splits into two peaks separated by 9 meV, very close to our experimentally determined faint gap (Fig. 2C). This feature arises as a consequence of the Coulomb repulsion. The end state splits at $U=0.8$ eV into a singly occupied and a singly unoccupied orbital, while the low-energy LUMOs are shifted to more negative energies, mixing with each other in the range of $U\sim1$ eV and forming two molecular orbitals that are hybrids of the non-interacting ones (see Supplemental Fig. S3). Fig. 2B shows the theoretical DoS distribution of the occupied state enclosed by the grey box in Fig. 2E, which is in excellent agreement with the experimental DoS in Fig. 2A.

Localized one-dimensional (1-D) states at the zigzag edges of graphene nanostructures\cite{nakada_edge_1996} have been predicted to become spin-polarized by MFH models\cite{yazyev_theory_2011,carvalho_edge_2014} and first-principles calculations. \cite{son_half-metallic_2006,hancock_generalized_2010,ortiz_engineering_2016,suda_energetics_2015} The driving mechanism is the energy gain of the system obtained by depleting a spin-degenerate doubly-occupied state near the Fermi level, for which the Coulomb repulsion would otherwise introduce a much higher internal energy. This phenomenon is prone to occur in localized electronic states, because the Coulomb repulsion energy grows as electrons get confined in a more reduced space.

Although our (3,1,8)-chGNRs do not have pure zigzag edges (see Fig. 1B), its periodic zigzag segments are known to retain intense localized edge states stemming from flat electronic bands near the $\Gamma$ point of the 1-D Brillouin zone (see Figs. 2A and 2D, and Supplemental Fig. S4B) \cite{sun_ch_from_2011,yazyev_theory_2011,suda_energetics_2015,li_topological_2021}.  If the two edges are far enough as to be considered independent, the edge states are expected to be metallic in (3,1,$c$)-GNRs\cite{yazyev_theory_2011}. However, in narrower ribbons, the coupling between both edge states can open a hybridization gap\cite{yazyev_theory_2011,li_topological_2021,merino-diez_unraveling_2018}. In the case of $c=8$, this gap amounts to approximately 20 meV, and is centered at the Fermi level\cite{li_topological_2021}. On (3,1,8)-chGNRs/GdAu$_2$, the slight electron doping causes an increase of the chemical potential in the ribbon, and instead of a gap at the Fermi level we find a partially filled state, at $\sim35$ mV above the HOMO (see Fig. 2C). This state is still close to the conduction band minimum (Supplementary Fig. S4), and therefore it will display a similar degree of localization at the edge as in the metallic case of wider ribbons ($c>8$)\cite{sun_ch_from_2011,yazyev_theory_2011,suda_energetics_2015}. This is illustrated in the aforementioned constant height scans with CO-tips (Fig. 2A and Supplemental Figs. S4 and S5) or in grids with current regulation for each pixel (Fig. 2D). Thus, they are excellent candidates to display magnetic instabilities associated to \emph{e-e} correlations.

The splitting induced by \emph{e-e} interactions endows different spin quantum numbers to the singly occupied/unoccupied states. From this, an inversion of the spin polarization at both sides of Fermi level follows necessarily.\cite{son_half-metallic_2006,yazyev_theory_2011,carvalho_edge_2014} In the following we provide evidence of such sign inversion in the spin polarization, setting the experimental hallmark for itinerant magnetism in edge states of nanographenes.

Figure 3A shows a $N=15$ chGNR, scanned under constant-height conditions for several magnetic $P$ or $AP$ states of the underlying GdAu$_2$ with homogeneous magnetization in the characterization region (see Supplementary Fig. S6 for a description of the magnetic history). The spin asymmetry in Fig. 3B, calculated from the $dI/dV$ maps at the energies of interest  as $S_a=100\times[(dI/dV)_{AP}-(dI/dV)_{P}]/[(dI/dV)_{AP}+(dI/dV)_{P}]$, is proportional to the spin polarization of the system right at the measurement energy. Figure 3D shows spectral lines along the chiral edges of the spin averaged DoS, i.e. $[dI/dV_{AP}+dI/dV_P]/2$. Here, the  previously discussed splitting for $N=21$ manifests again in the regions of larger DoS as two peaks at $V_b=-7$ mV and at $V_b=+5$ mV, noticeable at the bottom of the left edge (region 1 in Fig. 3A) and at the center of the right edge (region 3 in Fig. 3A). This intensity distribution results from the combined effects of the LUMO confinement pattern at the edge and the influence of the moiré periodicity (Supplemental Fig. S4 and S5).

Regions 1 and 3 are also the positions with clear spin contrast in Fig. 3B. Furthermore, we find experimentally a change of sign in the spin polarization across the Fermi level of $\pm8$ \%, in neat agreement with the predicted magnetic state driven by \emph{e-e} correlations. Therefore, the 12 mV gap observed around the Fermi level (Figs. 2C and 3D) can be safely attributed to a spin splitting that emerges to accommodate \emph{e-e} correlations in the partially occupied LUMO edge state. If these two peaks were to correspond to different quantum-well states of the conduction band, they would not have any different spin polarization than the inner region of the ribbon, and certainly their spin asymmetry would not be changing sign across the Fermi level. Note that the spin polarization of the underlying GdAu$_2$ obtained with the same kind of bulk Cr-tips is very small ($<4$ \%), and energy independent in this bias regime (Supplementary Fig. S2).

The energy dependence of the spin polarization is better visualized in single point $dI/dV$ spectra (Fig. 3E). As for the constant height images (Fig. 3B), the spin asymmetry is obtained as $S_a(eV_b)=(AP-P)/(AP+P)$, and represented by the green curves. All positions other than the ribbon periphery are characterized by featureless spectra whose spin asymmetry between $AP$ and $P$ states is below our experimental confidence ($\leq1.5$ \%). Region 1 and 3 exhibit the canonical behavior discussed above for a correlations splitting with $S_a=+7$ \% at $V_b=-10$ mV, and $S_a=-6$ \% at $V_b=+6$ mV. Region 2, with a much lower intensity of the edge state (see Fig. 3D), only displays a small signal of $S_a=-3.3$ \% at $V_b=+4$ mV, barely above our experimental uncertainty, and $S_a\simeq0$ at the energy of the negative bias peak.

Figure 3C displays the spin polarization obtained from MFH calculations with the set of parameters determined earlier ($U=1$ eV, 2 electrons added, see Methods), which is in good qualitative agreement with the experimental results (see Fig. 3B). This comparison allows us to understand the detected spin polarization. On the one hand, $U=1$ eV is smaller than the expected theoretical value for free standing nanoribbons, namely around 2-3 eV,  \cite{hancock_generalized_2010,schueler_optimal_2013,gunlycke_graphene_2007} which in our case is justified by the hybridization of the GNR $p_z$ orbitals with the metallic substrate. The low $U$ value induces a smaller splitting ($\sim$ 10 meV) than the FWHM of the molecular orbitals ($\sim$ 30 meV, see Fig. 2C), leading to the weakening of the spin polarization. On the other hand, the spin polarization of one edge is not homogeneous, contrary to the expectation for the ground state of isolated nanoribbons. This is ascribed to the hybrid character of the edge states, a consequence of electron doping and the orbital mixing associated to the $U$-term (Supplemental Fig. S3). Both of them favour the concentration of unoccupied quantum well states of the charge-neutral ribbon towards the Fermi level. As a result, the edge spin polarization displays oscillations (Figs. 3B-C).

The spatial integration of the spin polarization in Fig. 3B yields a non-zero total spin for occupied states. Furthermore, the magnetization of both edges does not appear fully correlated as in the theoretical simulation shown in Fig. 3C.
As discussed in Supplementery Note 3, the prevailing mechanism that distorts the ideal magnetic ground state is the local exchange interaction with the substrate that, at $T=1.2$ K, overcomes the antiferromagnetic coupling between the edges\cite{son_energy_2006,carvalho_edge_2014,jung_theory_2009} --mediated by the overlap of the decaying edge states towards the interior--. To corroborate this idea, and using ribbons over GdAu$_2$ regions with inhomogeneous magnetization, we demonstrate the independent switching of the spin direction in a portion of the ribbon while the rest remains unchanged (Supplementary Fig. S8). This is controlled by the exchange coupling with the local magnetization of the GdAu$_2$, showing that this interaction is responsible for stabilizing the magnetic moment of the edge states against spin fluctuations and therefore, plays a fundamental role to gain access to the edge's spin polarization.

Altogether, by synthesizing chiral graphene nanoribbons on top of a magnetic GdAu$_2$ monolayer, we have been able to access with exquisite spatial and energy resolution the spin-polarization of its edge states by SP-STM/STS. Doing so, a long standing challenge has been resolved that not only reveals important differences with respect to the originally expected spin-polarization of charge neutral ribbons, but also sets the stage to similarly characterize the immense amount of magnetic carbon-based materials that are being synthesized lately.

%

\pagebreak

\section*{FIGURES}

\begin{figure}[!h]
\centering
\includegraphics[width=\columnwidth,keepaspectratio]{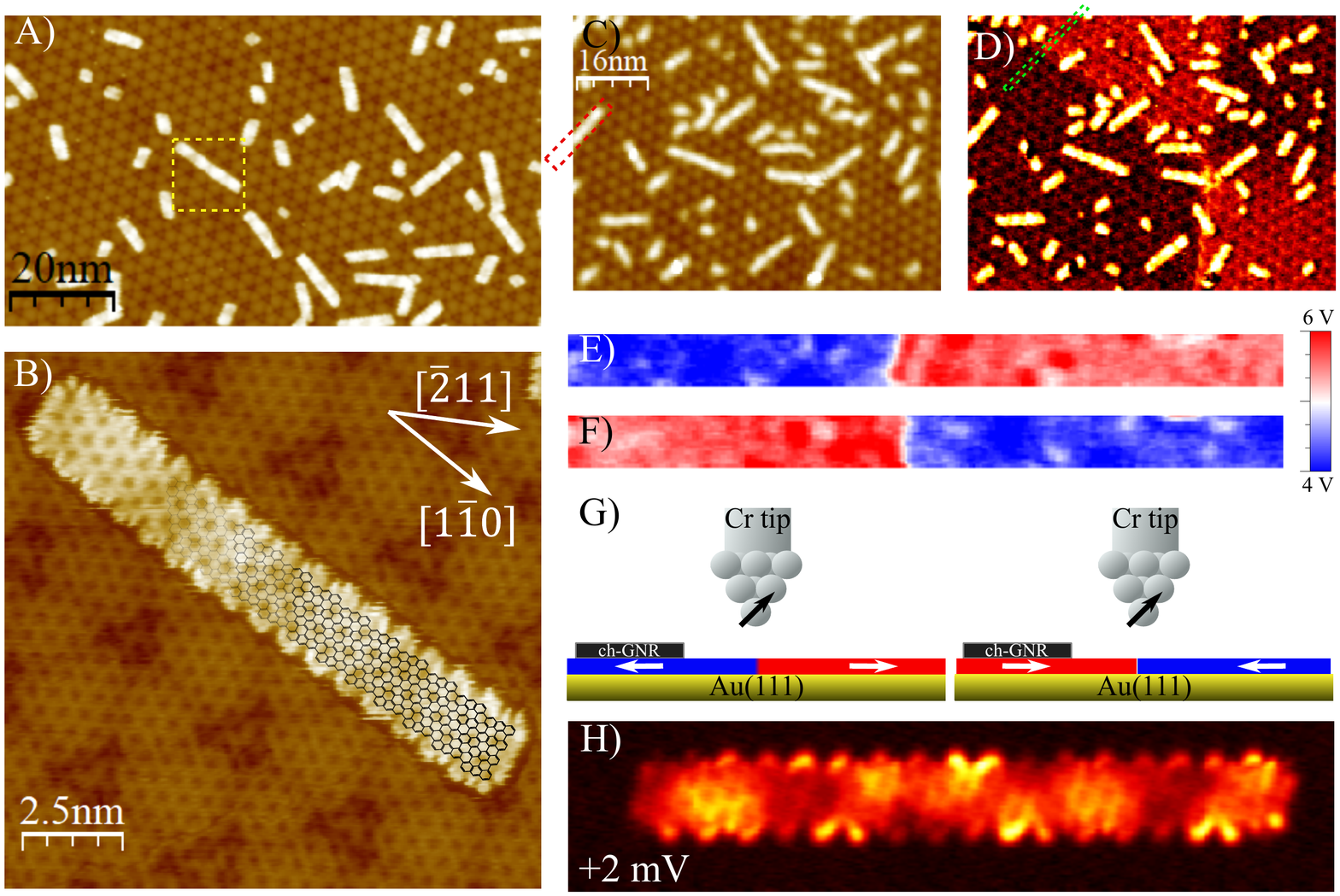}
\caption{\textbf{Experimental set-up for the magnetic chracterization of (3,1,8)-chGNRs on GdAu$_2$.} A) Survey of cGNRs on GdAu$_2$ (SP: -1.5V, 100 pA).B) Zoom of the area within the yellow square in A, STM topography mixed with the double derivative image to enhance the Gd lattice (SP: 1V, 50pA, W tip). C,D) Simultaneous topography image and dI/dV spin polarized map with in-plane sensitive Cr tip (SP: 3V, 50 pA; $V_{mod}$=20 mV rms; $B=0$ Tesla). The image contains structural antiphase boundaries (APB) which induce antiferromagnetic coupling among neighbouring domains. E,F) Zoom of the region enclosed by the green rectangle in D of two different remanent magnetic states ($B=0$ T) of the substrate obtained after cycling the field at maximum positive and negative out of plane field strength of $\pm3$ Tesla. This is an example of how we control the magnetic state of the tip-sample system for the subsequent magnetic characterization of the ribbons. G) Cartoon model of the entire tip-sample system in E and F where arrows represent the local magnetic moment. H) Constant height dI/dV maps at 2 mV ($V_{mod}=0.5$ mV rms) of the $N=17$ ribbon marked in C with the same Cr tip (feedback opened at ribbon center with SP: 50 mV, 300 pA).}
\label{Figure1}
\end{figure}

\newpage

\begin{figure}
\centering
\includegraphics[width=\columnwidth,keepaspectratio]{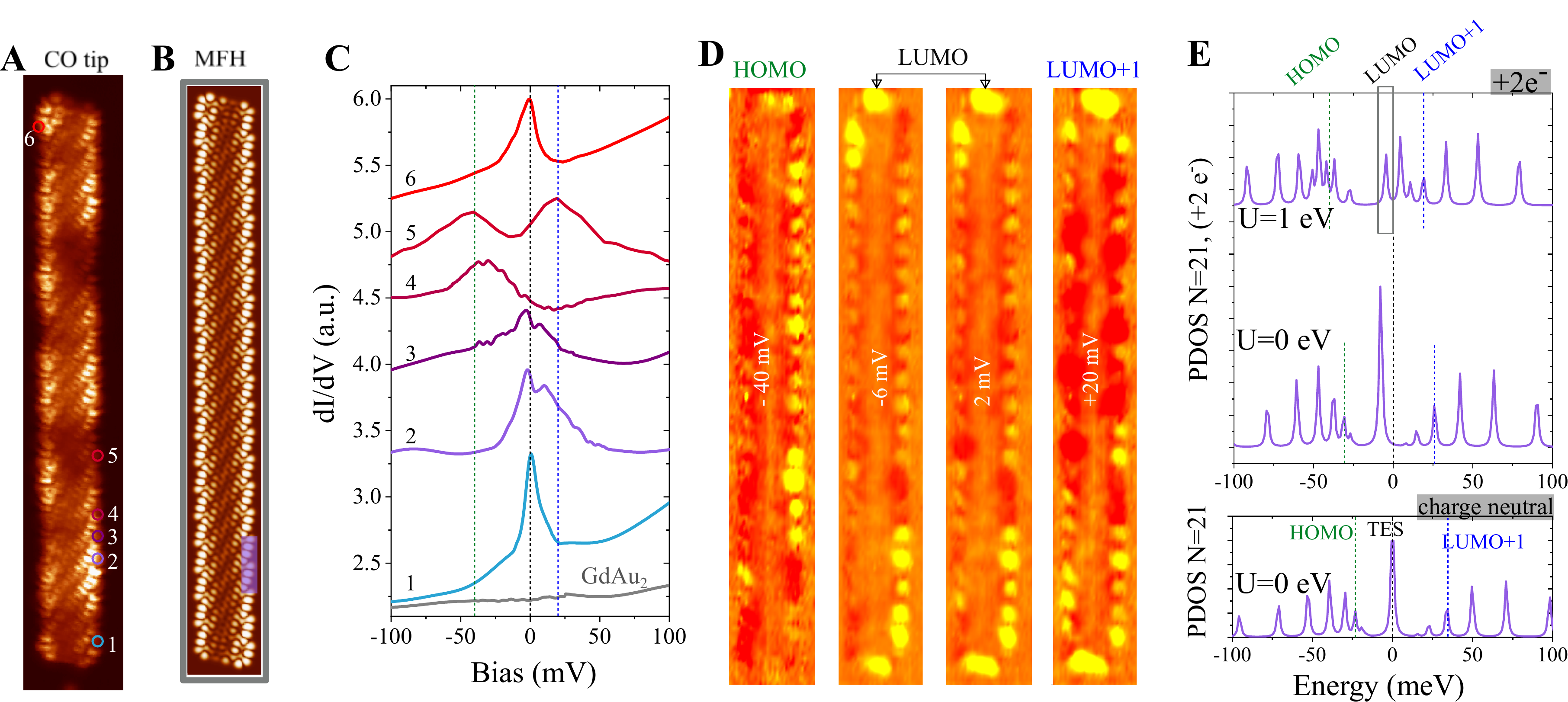}
\caption{\textbf{Spin averaged electronic structure of (3,1,8)-chGNRs on GdAu2.} ($T=4.3$K, $B=0$ T, W tip). A) Constant height current with CO functionalized W-tip at $V_b=-3.7$ mV of a $N=21$ ribbon (open feedback at ribbon's center with current SP of 50 pA). B) Simulation of the GNR DoS within the Mean Field Hubbard (MFH) model at the energy window enclosed by the grey box in E, $U=1$ eV and two electrond added (see also Methods). C) High resolution $dI/dV$ spectra  (SP: -100 mV, 200 pA, $V_{mod}$=0.7 mV) taken at the positions given by the colored circles in A. D) Constant current $dI/dV$ maps at the energy values marked by dashed lines in C with the same color code (SP: -150mV, 125pA, $V_{mod}=10$ mV). E) Projected DoS from MFH calculations (see Theoretical Methods) at the atomic sites enclosed by the transparent rectangle in B. Bottom panel corresponds to the charge-neutral case without \emph{e-e} interactions, top panel corresponds to the charged system with 2 electrons for $U=0$ and $U=1$ eV. Dashed lines in C and E indicate the energy positions of the HOMO (green), LUMO (black) and LUMO+1 (blue) named after their equivalent states in the charge-neutral specimen.}
\label{Figure2}
\end{figure}

\newpage

\begin{figure}
\centering
\includegraphics[width=\columnwidth,keepaspectratio]{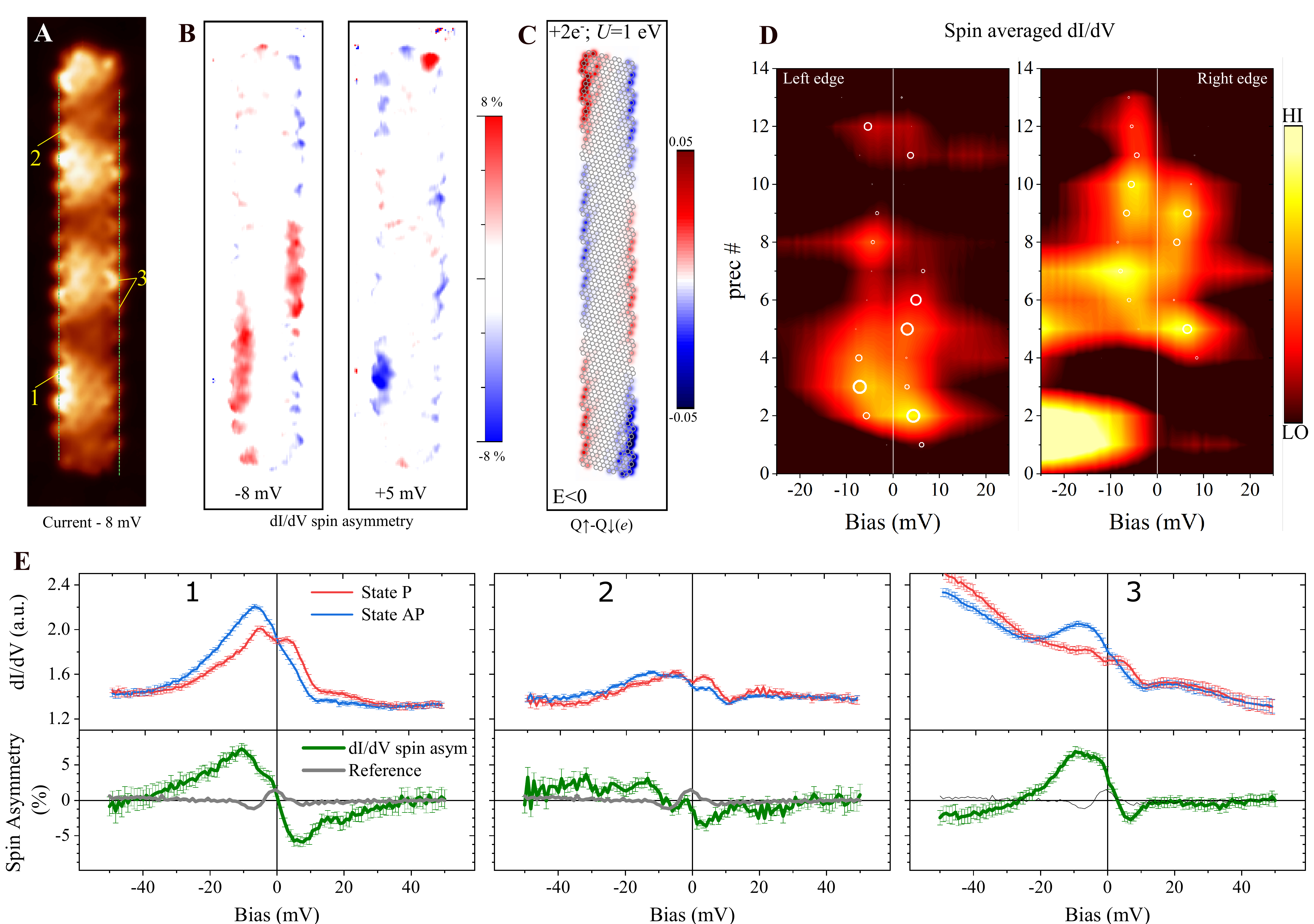}
\caption{ \textbf{Spin polarized edge states in chiral graphene nanoribbons.} ($T=1.2$ K, in-plane sensitive bulk Cr-tip, $N=15$ precursors). Control tests of the obtained magnetic contrast are provided at Supplementary Note 2. All images are $3.9\times15.8$ nm$^2$ and taken at constant height after opening the feedback over the ribbon’s center (SP: 20 mV and 100 pA, $V_{mod}=0.5$ mV). $dI/dV$ point spectroscopies taken with set point of 100 mV and 250 pA ($V_{mod}=0.5$ mV). A) Tunneling current image of the ribbon at $V_b=-8$ mV. B) $dI/dV$ spin asymmetry calculated from the constant height differential conductance images as $\% (AP-P)/(AP+P)$ at $V_b=-8$ mV (left) and $V_b=+5$ mV (right). Note the sign change of the spin polarization when crossing the Fermi level. C) Spin polarization of the GNR summed over occupied states, obtained with $U=1$ eV from the MFH model (see Methods). The spin polarization is spanned in a real space grid and the plot is obtained by slicing 3.5 \AA\ above the molecular plane. D) Stack plot of the spin averaged point spectroscopies, $(AP+P)/2$, taken along the green dashed lines in A. The position/diameter of the white circles represent the center/area of Gaussian functions used to fit each individual spectrum together with a baseline. E) Spin resolved point spectroscopy at the positions indicated by the yellow numbers in A for the $P$ (red curves,) and $AP$ (blue curves) states. The resulting energy resolved spin polarization is given by the green curves in the bottom panel of each graph. Light grey lines represent the background instrumental asymmetry obtained over GdAu2 or the ribbon center (both are identical). Error bars are the standard deviation with confidence of 68 \% obtained from 20 individual curves measured at each point.}
\label{Figure3}
\end{figure}

\clearpage



\section*{Acknowledgments}
We acknowledge financial support from the Spanish Ministry of Science and Innovation MICIN through grant nos. PID2019-107338RB-C64, PID2019-107338RB-C61, PID2019-107338RB-C62, PID2019-107338RB-C63, PID2020–115406GB-I00  funded by AEI/10.13039/501100011033; grant no. PCI2019-111933-2; and red tem\'{a}tica RED2018-102833-T. This work was also supported by  European Regional Development Fund (ERDF) under the program Interreg V-A España-Francia-Andorra (grant no. EFA194/16 TNSI), the European Union (EU) H2020 program through the FET-Open project SPRING (Grant Agreement No. 863098), the Maria de Maeztu Units of Excellence Program CEX2020-001038-M, the Aragon Government (E13-20R and E12-20R), the Programa Red Guipuzcoana de Ciencia, Tecnología e Innovación 2021 (Grant No. 2021-CIEN-000069-01. Gipuzkoa Next), the Basque Departmente of Educatioon (PRE-2021-2-0190 and PIBA-2020-1-0014), and the Xunta de Galicia (Centro de Investigación accreditation 2019–2022, ED431G2019/03)

\section*{Supplementary materials}
Experimental and Theoretical Methods\\
Supplementary Notes 1 to 3\\
Figs. S1 to S8\\
Supplementary References \textit{(46-55)}


\clearpage


\begin{center}
\LARGE{Supplementary information for:\\Detecting the spin-polarization of edge states in graphene nanoribbons}
\end{center}

\section*{Contents}

\begin{center}
\begin{tabular}{lr}
  Experimental Methods & 19 \\
  \\
  Theoretical Methods & 24 \\
  \\
  Supplementary Note 1.- Edge states distribution in real and reciprocal space & 26 \\
  \\
  Supplementary Note 2.- Monitoring the magnetic state of tip and substrate & 28 \\
  \\
  Supplementary Note 3.- Magnetic interactions with the substrate References & 33 
  
\end{tabular}
\end{center}


\section*{Experimental Methods}

\begin{figure}[h]
\includegraphics[width=0.9\columnwidth,keepaspectratio]{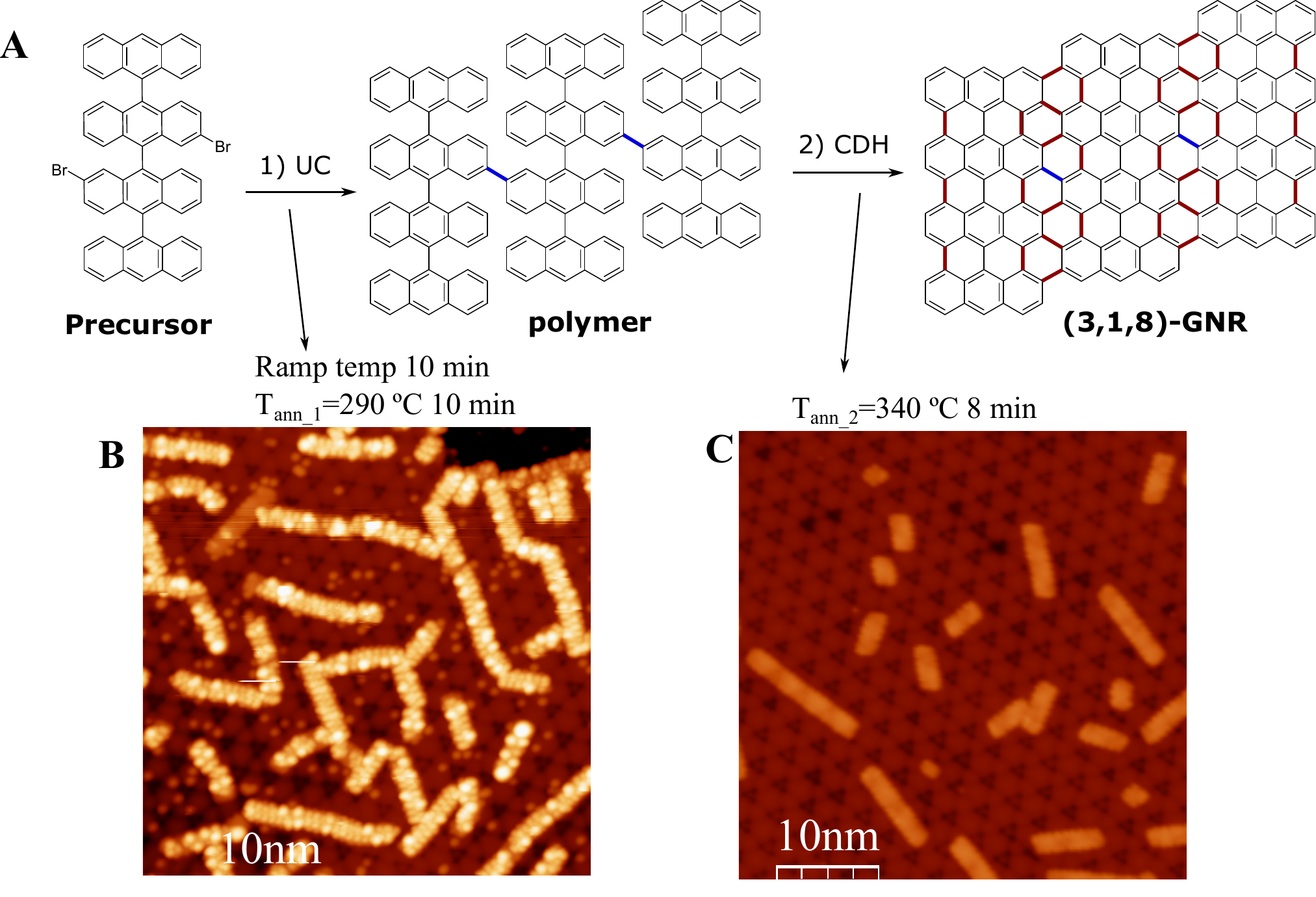}
\captionsetup{type=Supplementary Figure}
\caption{\label{fig:figS1} (\textbf{A}) Reaction scheme of the surface catalyzed synthesis of (3,1,8)-chGNRs starting from deposition of the reactant 2”,3’-dibromo-9,9’:10’,9”:10”,9”’-quateranthracene (DBQA, see Fig. \ref{fig:figS1}A), followed by Ullmann polymerization at approximately 290 $^\circ$C, and subsequent cyclodehydrogenation (CDH) at 340 $^\circ$C to achieve the final planar aromatic product. (\textbf{B}) Example survey of the polymeric chains of DBQA (set point -1.5 V, 100 pA). (\textbf{C}) Topography of fully planarized chGNR after CDH completion. B and C share the same color scale.}
\end{figure}

All measurements have been performed at the SPECS-JT-STM of the Laboratory for Advanced Microscopy (University of Zaragoza). This microscope attached to the sample preparation chambers, and features a base temperature of 1.17 K and an out-of-plane magnetic field up to 3 Tesla provided by a dry superconducting split-coil. The whole facility operates under ultra-high-vacuum conditions ($P\sim1\times10^{-10}$ mbar). The tip is grounded and the tunneling bias $V_b$ is applied to the sample. Data has been taken at $T=1.2$ K unless stated otherwise. Differential tunneling conductance $dI/dV$ is acquired using a lock-in amplifier at a frequency of 933 Hz and r.m.s. modulation given by $V_{\textrm{mod}}$. STM images and $dI/dV$ maps were taken either in constant height or in constant current mode, as indicated at the corresponding caption for each data set. In the case of the constant current $dI/dV$ maps as those shown in Fig. 2D of the main text and in Fig. S\ref{fig:figS4}B, the image is formed by slicing at fixed $V_b$ the $dI/dV$ signal from a dense grid where a full spectra is acquired ramping $V_b$ at each pixel. All images have been analyzed using WSxM software package\cite{horcas_wsxm_2007}.

\begin{figure}[ht]
\includegraphics[width=\columnwidth,keepaspectratio]{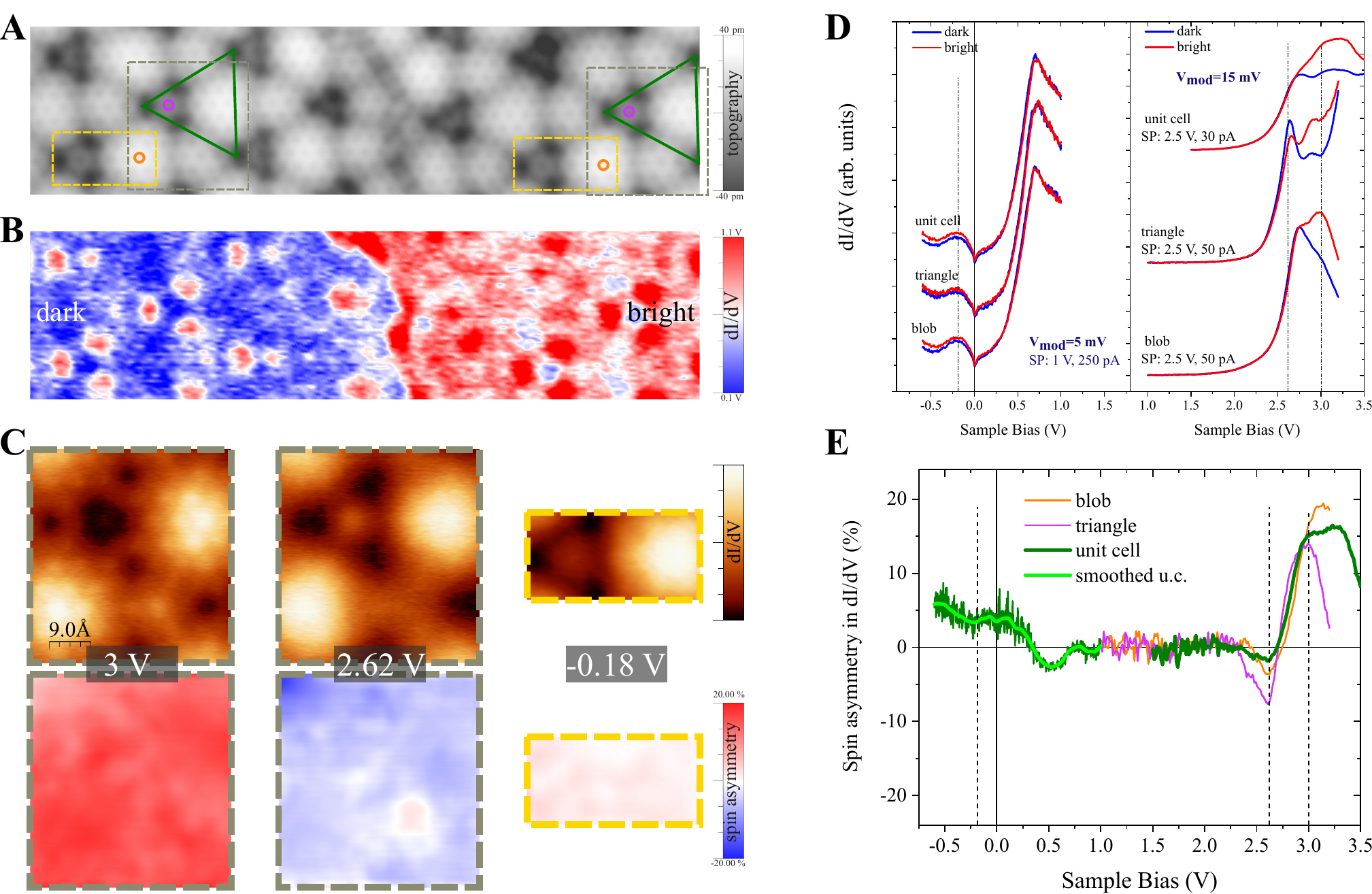}
\captionsetup{type=Supplementary Figure}
\caption{\label{fig:figS2} (Study of the spin polarization of GdAu$_2$ ($T=1.2$ K, in-plane sensitive bulk Cr-tip, $B=0$ Tesla). \textbf{A}) Atomically resolved STM topography (SP: 1 V, 500 pA) of a region with a structural antiphase boundary (APB) separating two crystallographic domains that are shifted by half a GdAu$_2$ lattice vector with respect to each other. \textbf{B}) Constant current spin resolved $dI/dV$ map of the same region (SP: 3 V, 50 pA; $V_\textrm{{mod}}=20$ mV rms) showing anti-parallel alignment of the magnetization across the APB. \textbf{C}) Constant height spin averaged (top row) and spin resolved (bottom row) $dI/dV$ maps obtained from the dashed grey and yellow rectangles marked in the domains with darker (D, left side in A and B) and brighter (B, right side in A and B) spin polarized conductance. In all cases, the spin averaged DoS map is then obtained as $(B+D)/2$, whereas the spin asymmetry map is obtained as $\%(B-D)/(B+D)$. Stabilization set points before opening the feedback on top of the magenta circles in A are 1 V at 35 pA for the maps at $V_b=$3 and 2.62 V ($V_{\textrm{mod}}=20$ mV rms), and 1 V at 1.5 nA for the map at $V_b=-0.18$ V ($V_{\textrm{mod}}=5$ mV rms). Spin asymmetry color scale spans $\pm20$ \%. \textbf{D}) Low-bias (left panel) and high-bias (right panel) single point $dI/dV$ spectroscopy obtained at the positions marked by the orange and magenta circles in A. The upper set of curves is an average over the approximate moir\'{e} unit cell depicted by the green triangles in A. Blue[red] curves correspond to the domains that are dark[bright] in B. Stabilization set points for the different spectra are given inside the graphs. \textbf{E}) Spin asymmetry calculated from the curves in D as $\%(B-D)/(B+D)$ for the following regions of the moiré pattern: center of the rounded blob (orange), center of the dark triangle (magenta) and approximate unite cell (green).}
\end{figure}

The Au(111) single crystal purchased from Mateck GmbH was cleaned by repeated Argon sputtering and annealing processes at 510 $^\circ$C. GdAu$_2$ alloy is grown on the clean Au(111) surface by sublimating Gd using an e-beam source at a rate of approximately 0.5 ML (referred to the Au(111) lattice) in 9 minutes while the substrate is held at 320 $^\circ$C. As shown in Fig. S\ref{fig:figS1}C, GdAu$_2$ forms a moir\'{e} superlattice caused by the superposition of its hexagonal unit cell (lattice parameter 5.41$\pm 0.03$ {\AA} along the $[\overline{2}11]_{Au}$ direction of the substrate) and the underlying Au(111) lattice\cite{corso2010_GdAu2,corso2010}. In the GdAu$_2$ lattice, each Gd atom is sixfold coordinated with Au atoms that are visualized as dark and bright spots respectively at $V_b=1$ V (see Fig. S\ref{fig:figS2}A and Fig. 1B of the article). The moir\'{e} pattern can be approximated by 4 GdAu$_2$ unit cells in the $[1\overline{1}0]_{Au}$ direction on 13 Au(111) lattice constants, with a period of $d_m=37.9\pm1$ \AA, although our own atomically resolved images indicate that the moir\'{e}e is not commensurate.

The reactant  2”,3’-dibromo-9,9’:10’,9”:10”,9”’-quateranthracene (DBQA) is then deposited on GdAu$_2$ from a home made resistive evaporator. Subsequent on-surface synthesis of (3,1,8)-chGNRs takes place in a two steps reaction as in the case of Au(111)\cite{li_topological_2021}, which is depicted in Supplementary Fig. S\ref{fig:figS1}. First, the temperature is slowly ramped up to 290 $^\circ$C and then mantained for 10 minutes to obtain long polymers via Ullmann coupling. Fig. S\ref{fig:figS1}B shows the characteristic corrugated topography of these kind of polymers, surrounded by Br atoms coming from the dehalogenation of the DBQA that are still on the surface. Second, the inner hydrogens of the polymer are cleaved by further heating up to 340 $^\circ$C to form the C-C bonds marked in red, resulting in the final fully planarized chGNR shown in Fig. S\ref{fig:figS1}C. Most of the GNRs grow with their longitudinal edge along high symmetry directions of the Gd atomic lattice ($[1\overline{1}0]_{Au}$ and $[\overline{2}11]_{Au}$ substrate directions), although there are as well about 20 \% of them oriented at 13$^\circ$ from these directions (the case of the marked ribbon in Fig. 1C main text), which corresponds to having the internal graphene lattice aligned with the high symmetry directions of the Gd mesh.

Spin polarized STM (SP-STM) has been carried out using bulk Cr tips. Tips are prepared by electrochemical etching of elongated pure Cr flakes and subsequent field emission cleaning (120 V, 1 $\mu$A, 1 hour) at the STM head. GdAu$_2$ exhibit atomically sharp structural antiphase boundaries (APB) between crystallographic domains that are shifted by half a GdAu$_2$ lattice vector with respect to each other, as the one shown in Fig. S\ref{fig:figS2}A. In agreement with a previous SP-STM study of GdAu$_2$\cite{bazarnik19}, we find that a vast majority of these APBs induce a local antiferromagnetic coupling of the neighboring domains, as evidenced by the spin polarized map in Fig. S\ref{fig:figS2}B. We make use of this contrast to calibrate the spin sensitivity direction and spin polarization of the Cr tips, as desccribed later in Supplementary Note 2. The tips are submitted to voltage pulses until the expected in-plane spin contrast of 20 \% or more at $V_b=3$ V shown in Fig. S\ref{fig:figS2}B is obtained.

$dI/dV$ point spectra at equivalent locations of the domains with high (bright, B) and low (dark, D) differential conductance are presented in Fig. S\ref{fig:figS2}D. They provide the quantitatively spin resolved electronic structure in the range of -0.5 to 3.5 eV around the Fermi level. In Fig. S\ref{fig:figS2}E the spin asymmetry is calculated from the $dI/dV$ spectra of representative moir\'{e} regions as $\%(B-D)/(B+D)$. Here it can be appreciated how the spin polarization at $V_b=3$ V is not homogeneous, but always positive and ranging 15 \% in average. At $V_b=2.6$ V there is an inversion of the spin polarization. These findings are further supported by the constant height maps of the electronic density and spin polarization shown in Fig. S\ref{fig:figS2}C. In contrast, the spin polarization around the Fermi level is much smaller, of about 4 \%. In addition, it is constant in a broad energy window of $\pm0.25$ eV and spatially homogeneous, as evidenced by the spin polarization map in Fig. S\ref{fig:figS2}C. This fact corroborates that the highly localized spin contrast and the spin polarization inversion discussed for chGNR's edge states in the main text are intrinsic to the ribbon, and not some kind of signal induced by the substrate. Note that, in addition, the ribbon plane probed by the tip in open feedback is $\sim$1.7 {\AA} above the substrate, and therefore with a much smaller contribution than the carbon atoms themselves.

\clearpage

\section*{Theoretical Methods}

To theoretically describe GNRs and their spin physics we employ the Hubbard Hamiltonian within the mean-field (MFH) approximation\cite{hubbard_electron_1964},
\begin{eqnarray}\label{eq:MFH-Hamiltonian}
H &=& -t_1 \sum_{\langle i,j\rangle,\sigma} c_{i\sigma}^\dagger c_{j\sigma}
-t_2 \sum_{\langle\!\langle i,j\rangle\!\rangle,\sigma} c_{i\sigma}^\dagger c_{j\sigma} -t_3 \sum_{\langle\!\langle\!\langle  i,j\rangle\!\rangle\!\rangle,\sigma} c_{i\sigma}^\dagger c_{j\sigma}\nonumber\\
&&+ U \sum_{i}\big(n_{i\uparrow}\langle n_{i\downarrow}\rangle
+\langle n_{i\uparrow}\rangle n_{i\downarrow}
-\langle n_{i\uparrow}\rangle \langle n_{i\downarrow}\rangle \big)
\end{eqnarray}
where $t_{1,2,3}$ are the hopping terms corresponding to the interaction between first, second and third nearest neighbors (3NN), respectively. For these parameters we use the numerical values $t_{1}=2.7$, $t_{2}=0.2$, $t_{3}=0.18$ eV \cite{hancock_generalized_2010}, corresponding to neighbor distances comprehended between $d_1<1.6$\AA$<d_2<2.6$\AA$<d_3<3.1$\AA, respectively.  For the Coulomb repulsion term, we use $U=1$ eV. The reason for using this 'small' value compared to other related works \cite{li_single_2019,li_uncovering_2020}, is because we see in this case a large interaction between the samples and the substrate, which leads to a screening of the electron-electron interactions.
Numerically, we solve the Schr\"odinger equation Eq. \ref{eq:MFH-Hamiltonian} using our custom implemented Python package \textsc{hubbard} \cite{dipc_hubbard}. Here, the average number operators $\langle n_{i\sigma} \rangle=\sum_{n}f_{\sigma, n}|b^n_{i\sigma}|^2$ for each spin component $\sigma=\lbrace \uparrow, \downarrow\rbrace$, are calculated by summing the eigenvectors ($b^{n}_{\sigma}$) resulting from the diagonalization of the MFH Hamiltonian, up to the last occupied $n$th molecular orbital (which depends on the present number of electrons), as encoded in the Fermi function $f_{n\sigma}$.

We compute the spin polarization for different $U$ values and electron energies, calculated as
\begin{equation}
\mathrm{Polarization}(i) = \sum_{n\sigma=\pm 1}\sigma |b_{\sigma ni}|^{2} f(E_{n\sigma})
\end{equation}
where $b_{\sigma ni}$ is the coefficient of the wave function projected on the 2$p_z$-orbital located at site $i$  for state $n$ and spin orientation $\sigma$ at energy $E_{n\sigma}$. $f(E_{n\sigma})$ is the Fermi function with $k_{B}T=10^{-5}$ eV.
We also plot this quantity spanned in a real space grid,
\begin{equation}
	\mathrm{Polarization}(\textbf{r}) = \sum_{n,i\sigma=\pm 1} \phi^{2}(\textbf{r}-\textbf{R}_i)\sigma |b_{\sigma ni}|^{2} f(E_{n\sigma}),
\end{equation}
where the carbon $2p_z$ basis orbital $\phi(\mathbf{r}-\mathbf{R}_i)$, centered at position $\mathbf{R}_i$, is described by the radial function $\exp{(-3|\mathbf{r}-\mathbf{R}_i|/\mathrm{\AA})}$.

Similarly, we obtain the LDOS as,
\begin{equation}
	\mathrm{LDOS}(\mathbf{r}, E) = \sum_n \mathrm{LDOS}_n(\mathbf{r})
	\frac{\gamma/\pi}{(E-E_n)^2 + \gamma^2},
\end{equation}
where
\begin{equation}
	\mathrm{LDOS}_n(\mathbf{r}) = \sum_\sigma \Big|\sum_i b_{ni\sigma} \; \phi(\mathbf{r}-\mathbf{R}_i) \Big|^2.
\end{equation}
The Lorentzian broadening, set to $\gamma=10$ meV for simulated spectra and  $\gamma=30$ meV for simulated images, is introduced to account for mix from energetically closely-spaced orbitals.

In all cases we slice the grid at a height of $z=3.5$ \AA\ above the chGNR. We plot these quantities using our \textsc{Hubbard} python package \cite{dipc_hubbard}.

\begin{figure}
\centering
\includegraphics[width=\columnwidth,keepaspectratio]{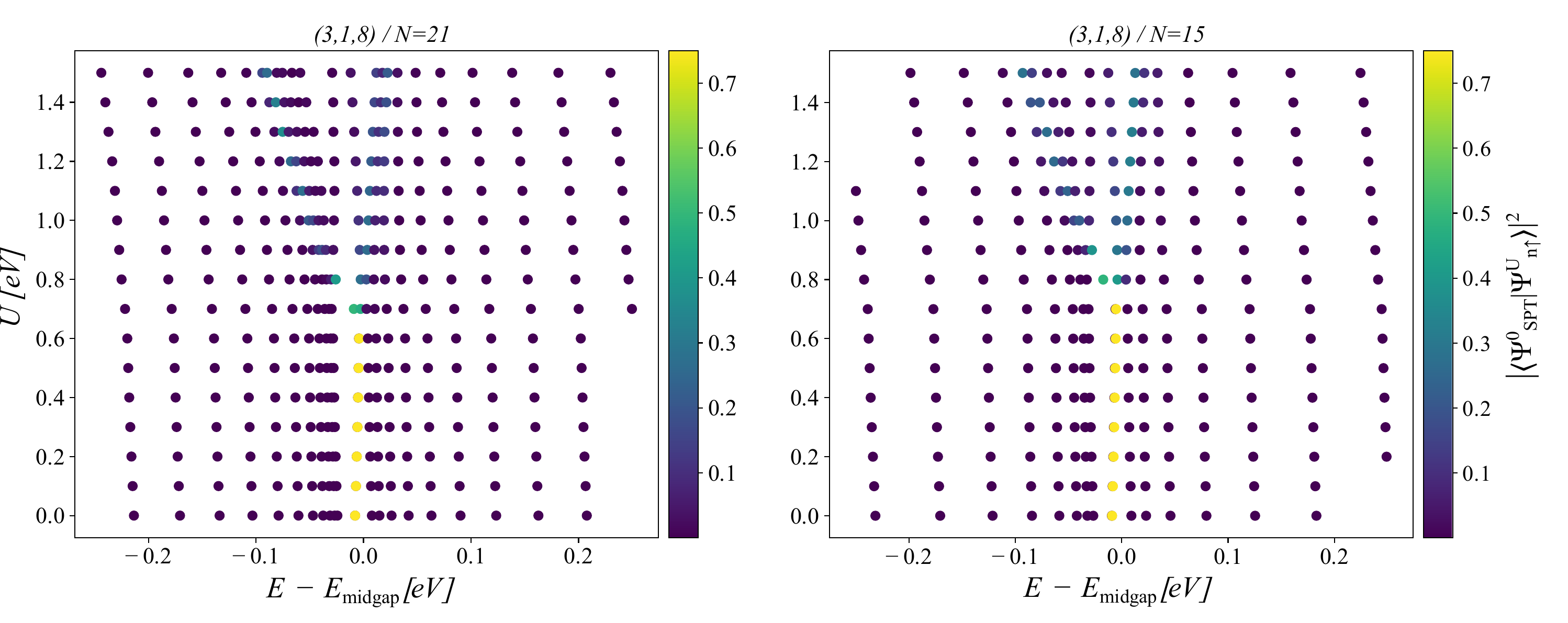}
\captionsetup{type=Supplementary Figure}
\caption{Charged system (2 electrons added): Overlap of the eigenstates of the MFH Hamiltonian (obtained with different $U$) for $\sigma=\uparrow$ with the SPT state obtained with $U=0$ (see eq. \ref{overlap}).}\label{fig:figS3}
\end{figure}

As a consequence of electron doping, the original zero-energy Symmetry Protected Topological (SPT) state\cite{li_topological_2021} of the neutral system is shifted to -8 meV after adding 2 electrons to the system (Fig. 2E main text). For increasing on-site Coulomb repulsion, this SPT state becomes hybridized with other orbitals at the following rate (color coded in Fig. S\ref{fig:figS3})

\begin{equation}\label{overlap}
\left|\langle \Psi^{0}_{\mathrm{SPT}}|\Psi^U_{n\uparrow}\rangle \right|^2
\end{equation}

where the $\Psi^{0}_{\mathrm{SPT}}$ state corresponds to the unoccupied SPT state and $\Psi^U_n$ in the $n$ eigenstate for varying $U$ after addition of 2 electrons to the charge neutral ribbon.

\section*{Supplementary Note 1.- Edge states distribution in real and reciprocal space.}

\begin{figure}
\includegraphics[width=0.9\columnwidth,keepaspectratio]{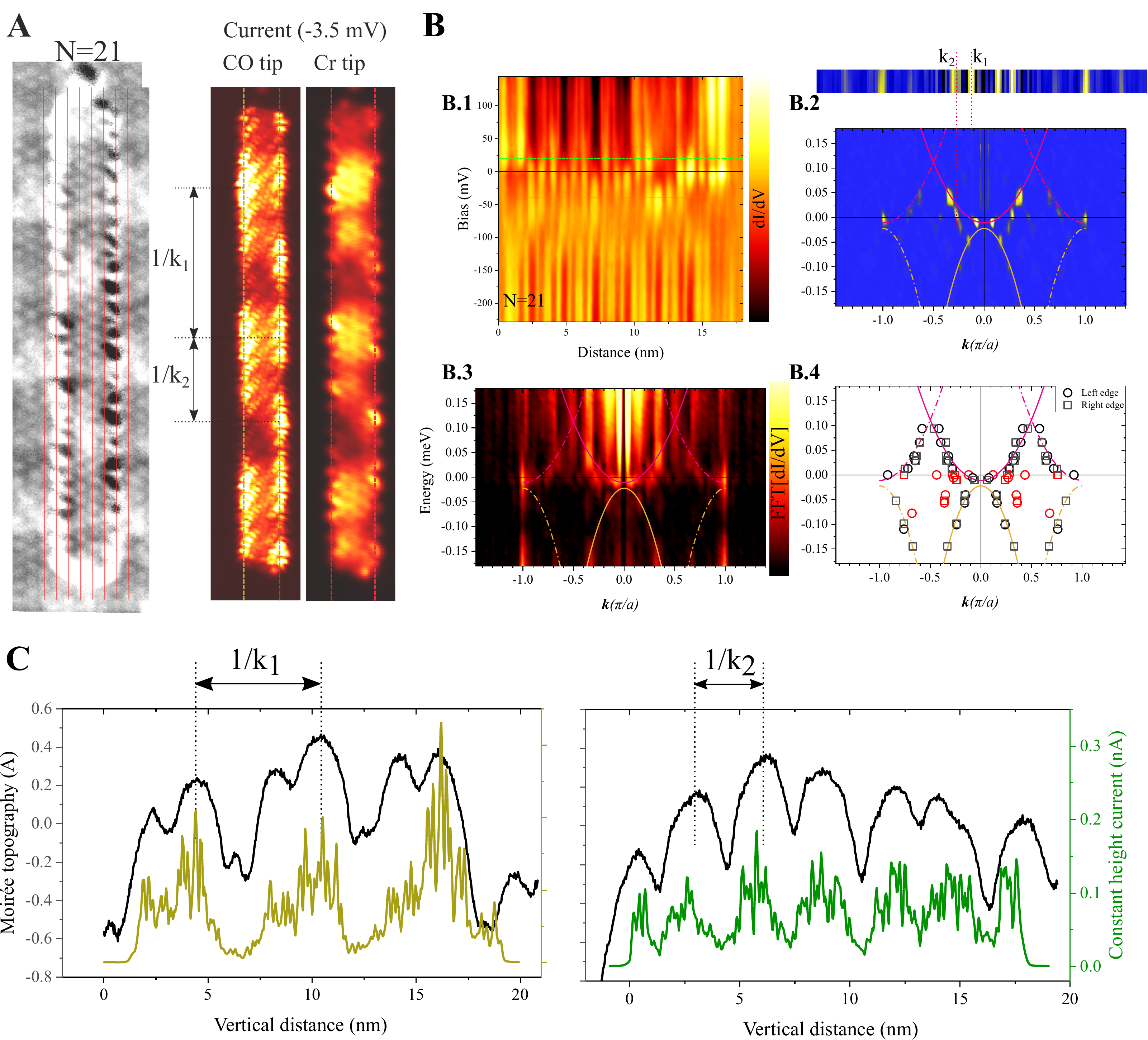}
\captionsetup{type=Supplementary Figure}
\caption{\label{fig:figS4} \footnotesize{A) Left panel is a $dI/dV$ \emph{constant current} map of the $N=21$ precursor units (3,1,8)-chGNR, oriented along $[\overline{2}11]_{Au}$ ($V_b=-100$ mV, SP: -100 mV, 200 pA, $V_{mod}=10$ mV), where Gd atoms are visualized as bright circles (thin red lines mark Gd rows). Right panel shows the \emph{constant height} tunneling current ($I_t$) of the same ribbon taken with a CO functionalized tip and a bare W tip (dull yellow and green dashed lines are the profile positions for panel C, red dashed lines mark the line spectroscopy position utilized for the analysis in panel B. B) $k$-space analysis of the periodicities in \emph{constant current} $dI/dV$ line spectroscopies taken along the red dashed lines in A (SP: 150 mmV, 125 pA, $V_{mod}=10$ mV, $a=8.87$ \AA\ is the 1D lattice parameter of the ribbon). B1 shows a stack plot of $dI/dV$ spectra as a function of the vertical distance in real space taken on the right edge. B2 is the line-wise Fourier transform (FT) of B1 after background subtraction of the raw transformed image B3. B2 and B3 are average FT of both edges. B4 is a collection of individual $E(k)$ spots obtained in the raw FT B3, and the pink and orange guidelines are best quadratic fits to the experimentally obtained $E(k)$. Red colored spots are excluded from the fit (see discussion in Supplementary Note 1). B2 also includes a 1D FT of the edge $I_t$ retrieved with the CO tip at -3.5 mV (average of both edges, profiles shown in panel C), where the characteristic moir\'{e} pattern frequencies ($k_1$ and $k_2$) are indicated by red dotted lines. C) Comparison of profiles of the moir\'{e} topography (SP: -500 mV, 500 pA) right below the ribbon's edges and of the constant height $I_t$ at -3.5 mV taken with the CO-tip along the dashed lines in A of the same color. Notice the correspondence between the edge brightness and the characteristic repetition patterns of the moir\'{e} along these lines ($1/k_1$ and $1/k_2$, which are also identified in B2).}}
\end{figure}

Figure S\ref{fig:figS4} contains the analysis of the electronic structure in reciprocal space of the chGNR with $N=21$ precursor units that is discussed in the main text. As shown in Fig. S\ref{fig:figS4}A, the edges (longitudinal axis) of this ribbon are parallel to the rows of Gd atoms (detected as bright spots in the $dI/dV$ map at -100 mV, panel A) running along the $[\overline{2}11]$ direction of the substrate. We define the ribbon's physical edge as the line where the tunneling current ($I_t$) experiences a sharp step down (dull yellow and green lines) in constant height mode using a CO functionalized tip. In this particular case, the edge lies between two Gd rows, i.e., on top of Au atoms. As shown in S\ref{fig:figS4}C, in this kind of constant height imaging, the edge state spatial distribution is heavily influenced by the moir\'{e} pattern of the underlying GdAu$_2$ monolayer. This is not related with the electronic structure of the edge states, but a consequence of the modulation of the local surface potential induced by the moir\'{e}, which gives rise to a periodically varying tunneling barrier. This effect can be partially suppressed working in constant current conditions, in which, for each pixel, the tunneling resistance is kept constant by regulating at a fixed set point of 150 mV and 125 pA. The resulting $dI/dV$ signal as a function of $V_b$ and longitudinal position is shown in Fig. S\ref{fig:figS4}B.1.

Line-wise Fast Fourier Transform (FFT) of energy and spatially resolved edge DoS (panel B.1) provides the 1D reciprocal space representation of the quasiparticle wave function modulus. In finite 1D ribbons, quantum confinement of the allowed $k$-vectors is expected to discretize the conduction and valence bands of the infinite counter parts, which have been reported elsewhere for (3,1,8)-chGNRs\cite{yazyev_theory_2011,li_topological_2021}. For long enough ribbons, a collection of allowed $k$-vectors and energies $E$ at which they appear, allows us the reconstruction of the dispersion relation, and thereby the identification of the band which the peaks of interest in spectroscopy (Figs. 2C and 3D main text) come from. This study is summarized in Figs. S\ref{fig:figS4}B.2-4. The $E(k)$ spots above 10 meV clearly follow a parabollic dispersion with positive effective mass (pink guidelines in S\ref{fig:figS4}B). For $0<E<10$ meV, two additional spots (at $k_1$ and $k_2$) appear, which we ascribe to the moir\'{e} periodicity discussed earlier (see Figs. S\ref{fig:figS4}B.2 and S\ref{fig:figS4}C). On top of them, a distinct intensity at $-10<E<-6$ meV with $k\neq k_1$ is identified, which we associate to the onset of the conduction band because it matches the quadratic fit to the $E(k)$ dispersion at higher energies. The next experimental spots are found at $E=$-40 meV, and together with those at lower energies, fit well another quadratic $E(k)$ dispersion with negative effective mass. As a consequence, we ascribe the peaks at $E\leq-40$ meV to confined states belonging to the valence band.

This analysis justifies the assignment of the observed peaks in Fig. 2C of the main text to HOMO (-40 meV), LUMO (Fermi level) and LUMO+1 ($\sim$20 meV) of the charge neutral ribbon. The straightforward conclusion is that the chGNRs are slightly electron doped on GdAu$_2$/Au(111). Owing to the sizable Coulomb repulsion ($U=1$ eV,  see eq. \eqref{eq:MFH-Hamiltonian}) required to describe precisely the DoS spectra (Fig. 2E main text), the confinement of edge states deviate from the trivial pattern of increasing integer number of nodal planes for increasing energy. This is neatly illustrated in Fig. S\ref{fig:figS3} for $N=21$ and $N=15$, for which the the non-interacting LUMOs heavily mix in the energy span of 20 meV around Fermi level. This effect is also reflected in the deviation of the experimental $E(k)$ from a free electron like quadratic dispersion, which is noticeable in Fig. S\ref{fig:figS4}B.4.

\begin{figure}[h!]
\includegraphics[width=0.775\columnwidth,keepaspectratio]{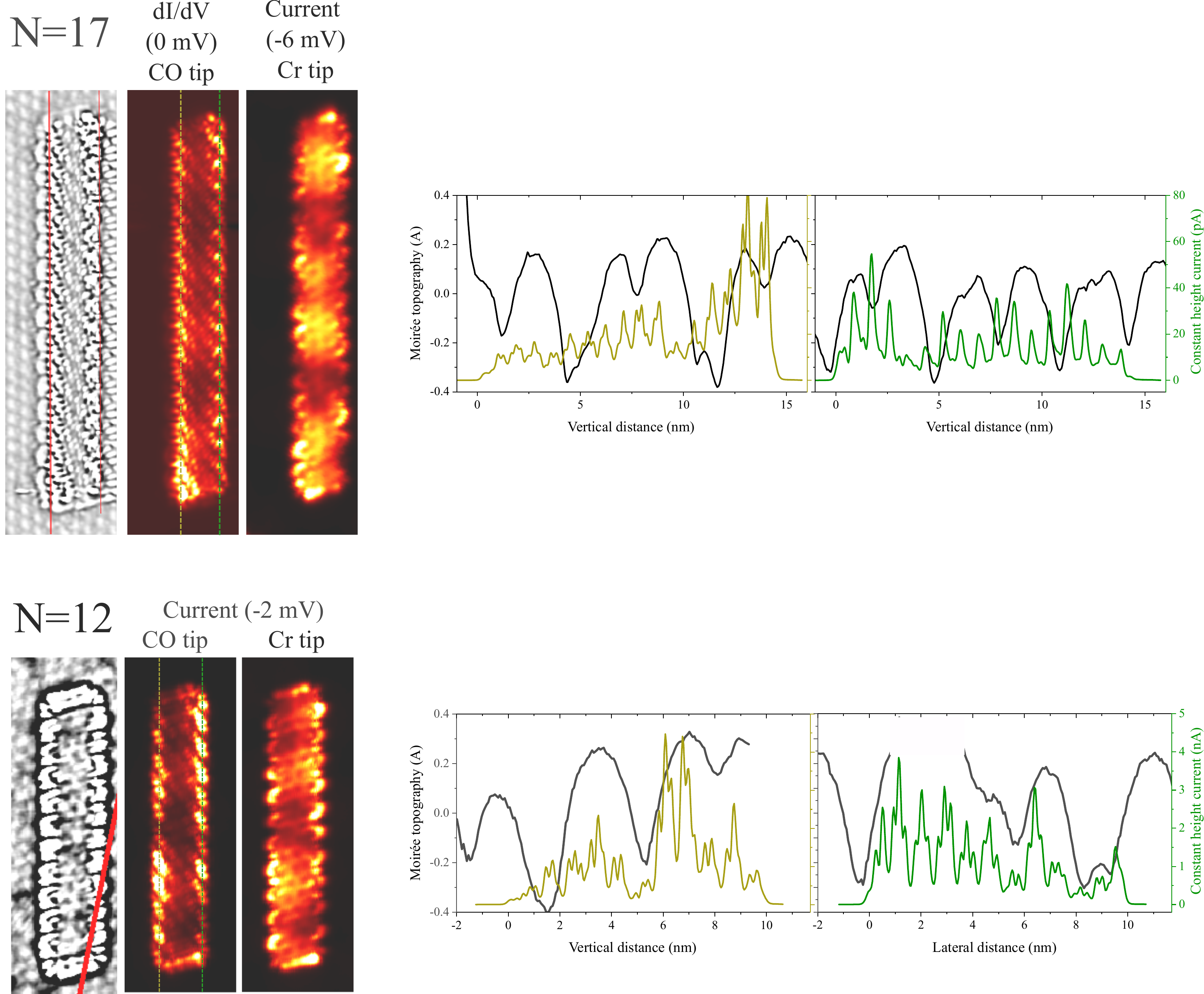}
\captionsetup{type=Supplementary Figure}
\caption{\label{fig:figS5} \footnotesize{Influence of the moir\'{e} pattern in constant height images of the $N=17$ and $N=12$ precursor units chGNRs with different stacking geometries over the substrate. In both cases, grey scale images are Laplace filtered STM topographies where the atomic Gd lattice and the ribbon perimeter can be visualized simultaneously (red lines mark Gd rows). The middle images are constant height DoS maps at Fermi level taken with a CO-functionalized tip Cr tip, while the right most images show the results for the bare metallic Cr tip. The profiles on the right panels display the moir\'{e} topography (SP: 20 mV, 500 pA) right below the ribbon's edges (black lines) and the constant height $I_t$ at very low bias taken with the CO-tip along the dashed lines of the same color in the corresponding image.}}
\end{figure}

Another remarkable experimental fact is that metallic tips fail to produce a clear DoS image of the chGNR eigenstates in constant height mode. When the $I_t$ image is retrieved by metallic tips, it portrays sizable contributions from the substrate (Gd lattice is visible in the center) and the edge state is not evident, as opposed to the case of using CO-tips or constant current mode with metallic tips (see Figs. S\ref{fig:figS4}A and S\ref{fig:figS5}). The CO-tip, from its part, tunnels efficiently to the chGNR states, because it guarantees the small tip-sample distance required to resolve them in constant height mode\cite{sode_electronic_2015}. For this latter case, the influence of the moir\'{e} pattern in the spatial distribution of the edge states discussed above is only negligible when the chGNR edge lies on top of Gd rows and parallel to the $[\overline{2}11]_{Au}$ direction ($N=17$ in Fig. S\ref{fig:figS5}). On the contrary, for edges along $[\overline{2}11]_{Au}$ over Au rows (Fig. S\ref{fig:figS4}A), or tilted with respect to that direction ($N=12$ in Fig. S\ref{fig:figS5}), the moir\'{e} pattern is replicated in the DoS profile.

\section*{Supplementary Note 2.- Monitoring the magnetic state of tip and substrate.}

\begin{figure}[!h]
\includegraphics[width=\columnwidth,keepaspectratio]{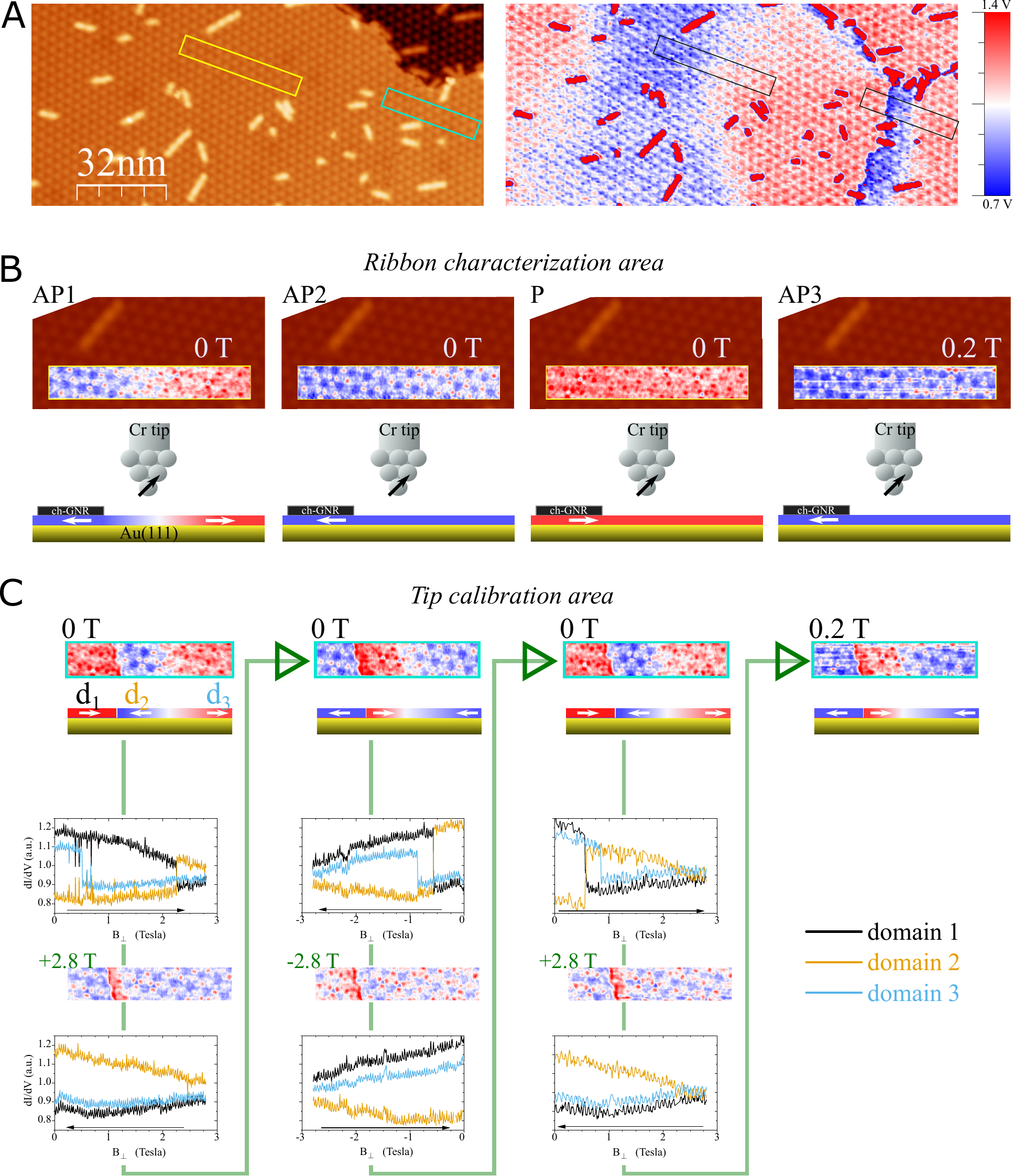}
\captionsetup{type=Supplementary Figure}
\caption{\label{fig:figS6} \footnotesize{Control of tip magnetic sensitivity during the acquisition of the data shown Fig. 3 main text ($V_b=3$ V, 50 pA, $V_{mod}=20$ mV, bulk Cr-tip). External field is perpendicular to the surface. The field variation history is indicated by the green arrows, and the field sweep direction by the black arrows in the graphs of panel C. The box size for RCA and TCA are $44\times7$ nm$^2$ and $35\times7$ nm$^2$ respectively.}}
\end{figure}

In order to uncover the magnetic moment distribution of the compelling chiral edge states, the SP-STM characterization has been performed with exquisite control of the magnetic state of the tip and the supporting substrate. One example of the procedure is given in Fig. S\ref{fig:figS6} concerning the $N=15$ chGNR discussed in Fig. 3 of the main text. The characterization starts by taking a spin resolved image of the chGNR environment in the virgin magnetic state (prior to applying external magnetic field), as shown in panel A. Next, a ribbon characterization area (RCA) and a tip calibration area (TCA) are chosen. RCA has to be close enough to the ribbon, but not necessarily including it to avoid over-scanning the primary sample. The RCA (yellow box) contains a control signal with the same spin dependent differential conductance as the GdAu$_2$ just beneath the chGNR. The TCA (green box) ought to contain one structural antiphase boundary (APB, see Fig. S\ref{fig:figS2}A,B) with antiferromagnetically coupled domains at either side. In this way, the TCA provides a well defined contrast to track the tip sensitivity direction and its spin polarization, which is used to guarantee that the tip's spin is unchanged throughout the whole process, incluiding the field ramps.

We use the following notation (see sketch in Fig. S\ref{fig:figS6}B) for the magnetic state of the RCA (and therefore that of the ribbon's region): $AP$ (or low $dI/dV$ signal at $V_b=3$ V, color coded blue) stands for the case in which the tip spin moment projection over the sample plane is mainly antiparallel to the RCA magnetization direction in the vicinity of the ribbon. $P$ (or high $dI/dV$ signal at $V_b=3$ V, color coded red) stands for the parallel case. The different states of the same magnetic character are numbered according to the chronological order in which they were set up. All changes of magnetic state in RCA and TCA are achieved by driving the external field up to $\pm2.8$ T and then back to zero. Taking into account the strong easy-plane magnetic anisotropy of GdAu$_2$ with coercivity of the order of 20 mT\cite{bazarnik19}, we attribute the field dependent contrast to the appearance of a small in-plane projection of the external field (caused by the unavoidable slight misalignement of the surface normal with respect to the vertical axis of the STM head). To understand the evolution of the magnetic contrast in the TCA, it is divided in three magnetic domains (see Fig. S\ref{fig:figS6}C). Domain 1 (black $dI/dV$ vs $B$ curves), to the left of TCA, is separated from domain 2 (brown curves) by an atomically sharp APB, and it is always antiferromagnetically coupled to domain 1. Domain 3 (blue curves), to the right of TCA, is separated from domain 2 by a natural magnetic domain wall approximately 9 nm wide.

The evolution of the RCA magnetic state and the unaltered spin moment of the probe tip can be appreciated in full detail in Figs. S\ref{fig:figS6}B-C. The original virgin state of the RCA is $AP1$. Sharp jumps as a function of field in the $dI/dV$ of the three domains correspond to local magnetization reversals caused by a fast moving domain walls crossing the measurement area. Note that the magnetization reversals occurring in domain 1 and domain 2 are always simultaneous and of opposite sign. This is due to an antiferromagnetic exchange interaction across the APB. Alternatively, this local contrast evolution could be also caused by a spin flip of the tip apex. However, we can unambiguously rule out any change in the tip's probe spin because, when this reversal takes place, the magnetization of domain 3 remains constant (pinned by other distant APBs). The saturation contrast of all three domains (see images at $\pm2.8$ T in panel C) is also illustrative in this respect. $dI/dV$ in all domains experiences a slow monotonous evolution with increasing $|B|$ towards an intermediate value between those of maximum contrast in remanence. This can be explained in terms of coherent rotation of the in-plane sample magnetization to become almost parallel to the field direction (while the tip spin sensitivity direction remains in-plane or slightly canted). In all field sweeps, the depinning field of the domain wall between 3-2 is larger than the depinning field of the wall traveling through domain 1 ($\sim\pm$0.85 T and $\sim\pm$0.5 T respectively).

\begin{figure}[!h]
\includegraphics[width=\columnwidth,keepaspectratio]{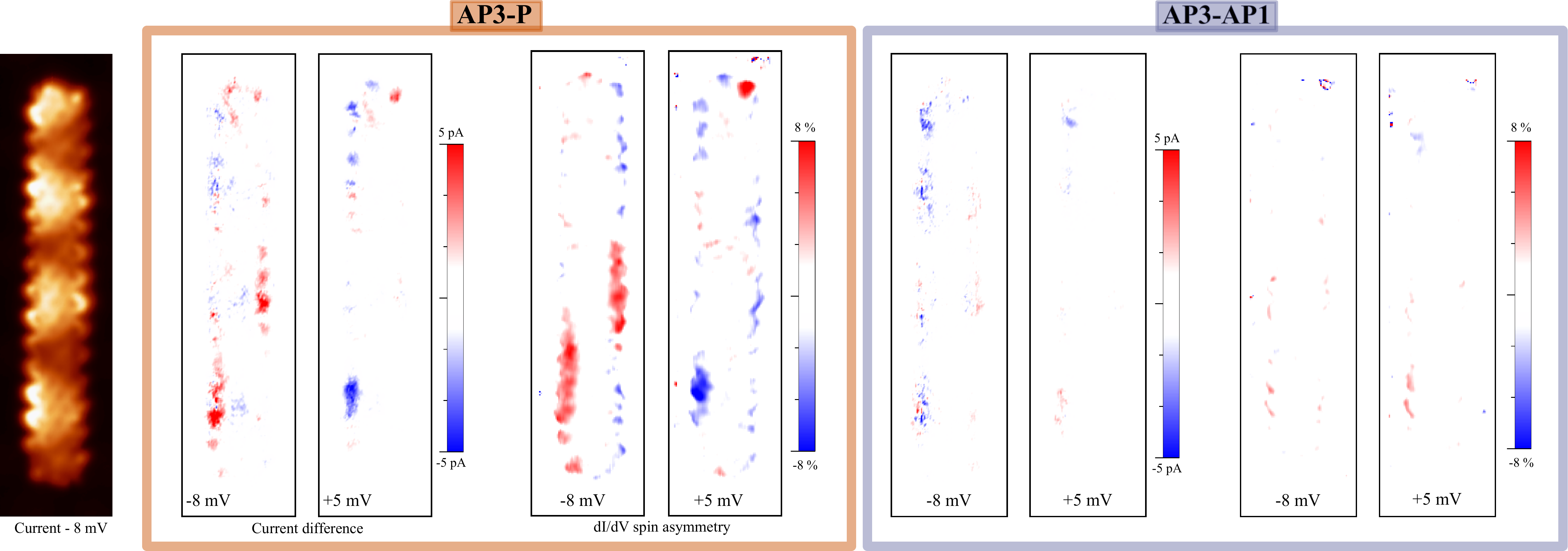}
\captionsetup{type=Supplementary Figure}
\caption{\label{fig:figS7} \footnotesize{Extended information about magnetic states in the experiment of Fig. 3, (in-plane sensitive bulk Cr-tip, $N=15$ precursors). All images are $3.9\times15.8$ nm$^2$ and taken at constant height after opening the feedback over the ribbon’s center (SP: 20 mV and 100 pA, $V_{mod}=0.5$ mV). The panel $AP3-P$ displays, at both sides of Fermi level, the magnetic contrast of the chGNR for two antiparallel magnetic states of the substrate. The direct current difference (left pair of images)  carries the energy integrated spin polarization from Fermi level to the measuring energy, whereas the spin asymmetry (right pair of images) is proportional to the spin polarization right at the measuring energy. The panel $AP3-AP1$ displays the direct current difference (left pair of images) and the spin asymmetry (right pair of images) for equivalent magnetic states of the substrate, resulting in the absence of spin contrast.}}
\end{figure}

The first magnetization ramp exhibits a somewhat different behaviour than the successive ones: during the sweep $AP1\rightarrow AP2$ the domain wall in RCA of the virgin state disappears, and the magnetization of the region between the ribbon an the TCA gets homogeneous, as evidenced in large area scans (not shown) and Fig. S\ref{fig:figS6}B. Afterwards, the magnetization of the RCA is homogeneous all the way to domain 1 and opposite to that of domain 2 in the TCA. In this way, the switch from $AP2\rightarrow P$ is induced by sweeping the field down to -2.8 T and then back to zero, and the switch from $P\rightarrow AP3$ by the same procedure but up to positive field of +2.8 T.

In order to extract the magnetic contrast discussed in the main text (Fig. 3) we have used $AP3$ and $P$ states. The result is reproduced in Fig. S\ref{fig:figS7}. Here, we plot the direct current difference $(AP3-P)$ and the $dI/dV$ spin asymmetry $(AP3-P)/(AP3+P)$. From this procedure, only the magnetic contrast associated to the change $P\rightarrow AP3$ remains in the signal. Identical results are obtained when $AP1$ and $P$ are compared. In contrast, when two images of equivalent magnetic states (like $AP3$ and $AP1$) are compared, there is not any intensity left in neither the current difference $AP3-AP1$ nor the $dI/dV$ spin asymmetry $(AP3-AP1)/(AP3+AP1)$. This is unequivocal proof of the magnetic origin of the contrast observed in chGNR edges among in-equivalent magnetic states of the supporting GdAu$_2$.

\section*{Supplementary Note 3.- Magnetic interactions with the substrate}

(3,1,8)-chGNRs belong to the class of graphene nanostructures with equal number of Carbon atoms in both sublattices of the honeycomb structure. Therefore, Ovchinnikov's rule\cite{Ovchinnikov_Multiplicity_1978} and Lieb's theorem\cite{lieb_two_1989} impose that the total spin is zero, although the spin density may be locally finite. In particular, it is predicted the antiferromagnetic (AFM) alignment of the spin-polarized edge states on either side of the ribbon. However, one important ingredient of this theorem is the particle-hole symmetry (half-filling). In our case, we deviate from this situation in two aspects: first, the GNR is slightly charge doped, and so its chemical potential shifts from the charge neutrality point; second, the Hubbard Hamiltonian that reproduces the electronic structure shown in Fig. 2 includes hopping terms up to third nearest neighbours (see eq.\eqref{eq:MFH-Hamiltonian} in Theoretical Methods). Under these conditions, the magnetization of opposite edges do not necessarily add up to zero and Lieb's theorem does not hold\cite{carvalho_edge_2014, Sawada_Phase_2009, Jung_Carrier_2009}. Both deviations are small; the filling amounts to $2/21\simeq0.1$ electrons per unit cell (or $\sim0.03$ electrons per zigzag lattice vector), while the ratio of second and third neighbors hopping relative to nearest neighbors is $\sim0.07$. However, it is enough to justify a finite but small spin-moment of the overall ribbon caused by the deviation from the expected AFM alignment between edges\cite{Jung_Carrier_2009,Sawada_Phase_2009}. Note that the long standing prediction of AFM coupling between edges was obtained in \textit{ab-initio} and model Hamiltonian calculations only for the case of charge neutral ribbons. On the contrary, in zigzag GNRs with the same witdh as the (3,1,8)-chGNRs and small electron doping comparable to our case, MFH calculations find magnetic solutions where the edge magnetizations are of opposite sign but not of the same value\cite{Jung_Carrier_2009}.

The intrinsic spin moment of the chGNR could couple with the external magnetic field or undergo exchange interactions with the substrate. To rule out the influence of the external magnetic field we performed all measurements at very small magnetic fields ($B<200$ mT), exploiting opposite remanent states of the substrate (see Fig. 1). Zeeman energy at these fields is orders of magnitude smaller than the thermal fluctuations at the experimental temperature of $\sim1$ K ($0.08$ meV).

\begin{figure}[h!!]
\centering
\includegraphics[width=\columnwidth,keepaspectratio]{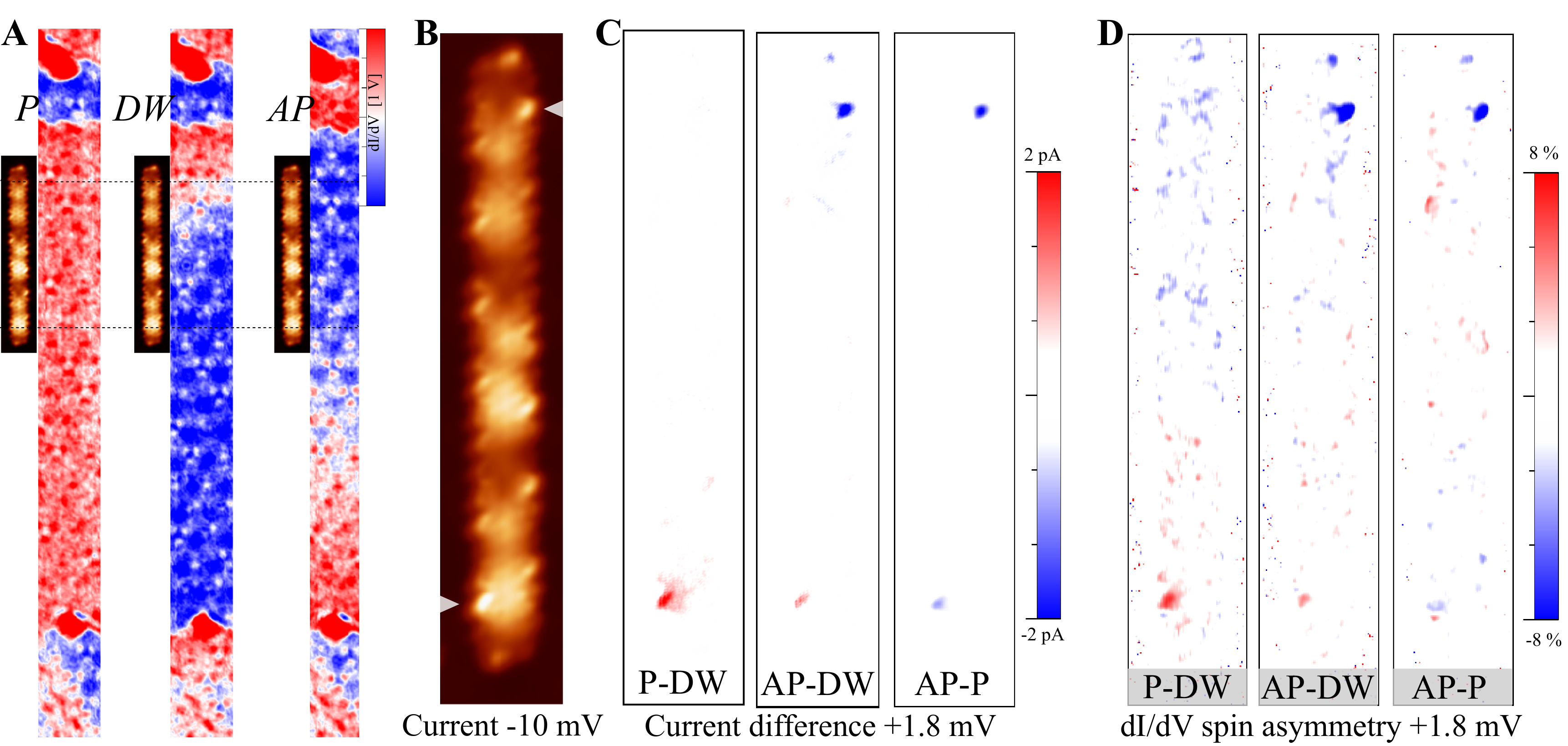}
\captionsetup{type=Supplementary Figure}
\caption{\footnotesize{\textbf{Magnetic interaction with the substrate.} Spin resolved DOS of a N=22 (3,1,8)-chGNR on GdAu2 (T=1.2 K, $B=0$ T, bulk in-plane sensitive Cr tip) across a magnetic domain wall. A) SP-STM $dI/dV$ map (3V, 50 pA, $V_{mod}=20$ mV, vertical length 80 nm) of the GdAu2 region immediately to the right of the ribbon (the small tunneling current image next to each map represents the true position of the ribbon) in three different magnetic states of the substrate obtained after different field cycles: $P$ ($B=0$ Tesla, minor cycle up to 2.3 T) exhibits in-plane magnetization parallel to the tip’s magnetic moment in the region underneath the ribbon, $AP$ ($B=0$ Tesla, full cycle up to 3 T) shows mainly antiparallel alignment, while in the $DW$ state ($B=0$ T, full cycle up to -3 T) a natural domain wall about 10 nm wide has formed. B) Tunneling current image of the N=22 (3,1,8)-cGNR at $V_b=-10$ mV. Triangles are placed at the same position as the dashed lines in A marking the spots where a finite spin asymmetry of the edge state has been detected. C) Direct current difference at $V_b=1.8$ mV between magnetic states $P/AP$ and $DW$. D) $dI/dV$ spin asymmetry ($S_a$) calculated from the constant height differential conductance images in $P$,$AP$ and $DW$ states as $100\times(P[AP]-DW)/(P[AP]+DW)$ at $V_b=1.8$ mV ($V_{mod}=0.5$ mV). All images in B,C,D (4$\times$22 nm2) are taken constant height at the tip sample distance corresponding to a set point of 20 mV and 100 pA at the ribbon’s center. In C,D positive $S_a$ is observed in the bottom part of the ribbon only when comparing $P$ and $DW$ (see triangles in B), which is the position where $P$ and $DW$ states shows strong contrast in the GdAu$_2$ below the ribbon (see dashed lines in A). In the case of subtracting the images in $AP$ state from the ones in $DW$ state, we have the same result in the upper part of the ribbon, which is now the only region of the substrate with sizable magnetic contrast. For the $AP-P$ case, for which the whole ribbon is approximately over GdAu$_2$ with opposite magnetic moment, $S_a$ in the top and bottom part behave as in the top part for $AP-DW$ case.  In the spin averaged point spectra, these two spots are the only locations with strong intensity at Fermi level, suggesting that in the rest of the edge no traces of spin polarization are expected.}}\label{fig:figS8}
\end{figure}

In our system, there is another possible magnetic interaction: the exchange coupling with the substrate's magnetization. To investigate its influence on the spin distribution of the supported chGNRs, we designed an experiment involving an intermediate state with in-homogeneous magnetization under the ribbon (i.e., a natural domain wall of the GdAu$_2$ across the ribbon is formed) , which is summarized in Fig. S\ref{fig:figS8}. Here, as shown in panel A for a $N=22$ GNR, the magnetization of the GdAu$_2$ between $P$ and $AP$ states is opposite underneath the ribbon, but in the domain wall ($DW$) state the top part of the ribbon lies in a region with the same magnetization as the $P$ state, whereas the bottom part lies in a region with the same magnetization as the $AP$ state. In this way, we detect a small spin asymmetry $S_a>0$ at Fermi level extracted from the difference $(P-DW)$ only in the bottom part of the left edge (Figs S\ref{fig:figS8}C,D). For the $(AP-DW)$ case, $S_a$ is zero in the bottom part, and $S_a<0$ in a small spot of the right edge at the top. Thereby, it can be concluded that the spin polarization does not behave like a strongly correlated spin density of the overall molecule, but it rather couples \emph{locally} to the magnetic moments of the substrate. Uncorrelated inter-edge spin configurations are those for which there is no energy difference between AFM or ferromagnetic alignment. In the case of zigzag GNRs with $c=8$ the energy gain of the AFM arrangement has been estimated to be $\sim2$ meV\cite{jung_theory_2009,lee_magnetic_2005} per edge atom. The exchange interaction in chGNRs is even weaker\cite{yazyev_theory_2011}, so this number is just an upper threshold. At the same time, although all SP-STM data discussed so far have been taken at $T=1.2$K, we have obtained a similar $S_a$ distribution to the one shown in Fig. S\ref{fig:figS8} at $T=4.4$ K for a $N=26$ (3,1,8)-chGNR. This suggest that the average exchange interaction between carbon atoms and the substrate must be significantly higher than the thermal energy of $\sim$0.4 meV, which leads to the conclusion that it determines the inter-edge spin alignment at a greater extent than small perturbations of the already weak internal AFM coupling.

More importantly to our work, this interaction with the substrate is responsible for stabilizing the magnetic moment of the edge states against thermal\cite{meier08} and quantum fluctuations\cite{donati_magnetic_2016}. Available estimates of the spin-orbit coupling strength in graphene provide the figure of 15 $\mu$eV\cite{Slota_Magnetic_2018}, which in a uniaxial approximation for the magnetic anisotropy would provide a energy barrier of the order of only 0.2 K. This means that the thermal fluctuations at any temperature near 0.2 K would suffice to destabilize spin moments in nanographenes, or in other words, the spin correlation lengths would be smaller than the characteristic dimensions of the ribbbons\cite{yazyev_magnetic_2008}. In our experimental approach, however, this is not happening thanks to the exchange coupling to the GdAu$_2$. For ribbons positioned over homogeneous magnetization regions, as corresponds to the model case discussed in the main text, this magnetic interaction plays, therefore, a fundamental role to gain access to the edge's spin polarization. It enables the access to the spin density by means of slow SP-STM scans under different configurations of the tip-sample magnetization, and the demonstration of magnetic remanence in the extended 1-D edge states of GNRs with high density of zig-zag segments.

\clearpage




\end{document}